\documentclass[aps,prl,amsmath,amssymb,twocolumn,superscriptaddress]{revtex4}
\usepackage[english]{babel}
\usepackage{graphicx,bm,amssymb,amsmath}
\usepackage{color,wrapfig,braket}
\usepackage{hyperref,placeins,ulem}
\usepackage[caption=false]{subfig}
\usepackage{subfig}
\usepackage{ulem}

\usepackage{epsfig}
\usepackage{verbatim}
\usepackage{yhmath}
\def\S{\textrm{S}}

\def\HG{\mathcal{H}_G}
\def\distAB{\text{dist}(A,B)}
\def\Tr{\text{Tr}}
\def\ShannonTO{4}

\makeatletter
\newcommand\xleftrightarrow[2][]{%
  \ext@arrow 9999{\longleftrightarrowfill@}{#1}{#2}}
\newcommand\longleftrightarrowfill@{%
  \arrowfill@\leftarrow\relbar\rightarrow}
\makeatother

\begin{document}

\title{Long-range mutual information and topological uncertainty principle}
\author{Chao-Ming Jian}
\affiliation{Department of Physics, Stanford University, Stanford, California 94305, USA}
\author{Isaac H. Kim}
\affiliation{Perimeter Institute for Theoretical Physics, Waterloo, Ontario, Canada}
\affiliation{Institute for Quantum Computing, University of Waterloo, Waterloo ON N2L 3G1, Canada}
\author{Xiao-Liang Qi}
\affiliation{Department of Physics, Stanford University, Stanford, California 94305, USA}
\date{\today}

\begin{abstract}
Ordered phases in Landau paradigm can be diagnosed by a local order parameter, whereas topologically ordered phases cannot be detected in such a way. 
In this paper, we propose long-range mutual information(LRMI) as a unified diagnostic for both conventional long-range order and topological order. Using the LRMI, we characterize orders in $n+1$D gapped systems as $m$-membrane condensates with $ 0 \leq m \leq n-1$. The familiar conventional order and 2+1D topological orders are respectively identified as $0$-membrane and $1$-membrane condensates. We propose and study the topological uncertainty principle, which describes the non-commuting nature of non-local order parameters in topological orders.
\end{abstract}

\maketitle
{\bf Introduction} - In the Landau paradigm, conventional orders (CO) are characterized by the phenomenon of spontaneous symmetry breaking. In the thermodynamic limit, some symmetry-breaking local operator--the order parameter-- attains an expectation value. Alternatively, one can also consider the symmetry-preserving ground state and describe CO by a nonvanishing long-range correlation between order parameters, as was proposed in the case of off-diagonal long-range order\cite{Yang1962}. Another kind of order that is beyond the Landau paradigm, the topological order (TO), has been studied since the discovery of fractional quantum Hall effect\cite{Tsui1982}. TO's are usually characterized by different theoretical frameworks such as topological ground state degeneracy or braiding statistics of the topological quasi-particles. While non-local order parameters have been proposed in some particular TO's such as Laughlin state\cite{Zhang1992} and $Z_2$ spin liquid \cite{Balents1999,LevinSenthil2004},
much less is known about such order parameters in general TO's.
Since systems with CO and TO both have robust ground state degeneracy in the thermodynamic limit, it is natural to ask whether they can be described in a unified framework. Such a framework is likely to provide new insight on the definition and characterization of TO in higher dimensions, which is much less understood than that in two spatial dimensions.

In this letter, we propose a new measure of long-range entanglement, 
the long-range mutual information (LRMI), as a unified language to describe both CO and TO. Mutual information between two regions of a many-body system is known to upper bound all connected correlation functions them\cite{Wolf2008}. Therefore, the LRMI between far away regions with different topology should be a good indicator of both CO and TO. In the following, we will provide the definition of the LRMI. Then we shall study the LRMI diagnostic of the CO and different TOs in 2+1D, as well as in higher dimensions. As it shall become clear, we view the CO as a condensate of particles, 2+1D TO as a condensate of strings, and higher dimensional TO as that of extended objects of various dimensions. Therefore, the topology of the chosen regions plays an important role in understanding the nature of the underlying order. 
The LRMI description also explains the key difference between TO and CO. The non-local order parameters in TO can be defined on intersecting regions, which generically do not commute with each other. The non-commuting nature is characterized by a ``topological uncertainty principle", which states that a pair of LRMI on these intersecting regions cannot vanish simultaneously. The topological uncertainty principle is a smoking-gun indication of long-range quantum entanglement in TO. While we will mainly focus on gapped systems with a finite correlation length $\xi$ in this letter, we shall also provide some comments on gapless systems as well.

{\bf Definition of the LRMI} - 
For a given quantum many-body state, we denote $\rho_A$ as the reduced density matrix of a region $A$. The entanglement entropy of $A$ is given by $S_A= S(\rho_A) \equiv - \text{Tr} \rho_A \log \rho_A$. For two disjoint regions $A$ and $B$, their mutual information is defined as
\begin{align}
I_{A,B} \equiv S_A + S_B - S_{A \cup B}.
\end{align}
For our purpose, we always take regions $A$ and $B$ to have the same topology (denoted as $\mathcal{X}$) and take the limit in which the distance between $A$ and $B$($\text{dist}(A,B)$) is infinite. In this limit, we denote the asymptotic value of $I_{A,B}$ as $\mathcal{I}(\mathcal{X})$, and refer to it as the LRMI\footnote{It goes without saying that the long distance limit is only well-defined when the system size is taken to infinity and $A$ and $B$ do not intersect with each other.}. There are different choices of $\mathcal{X}$ that characterize different types of orders. For example, on a two-dimensional torus, $\mathcal{X}$ can be a disk or a cylinder. As we shall see later, these choices work as a probe to detect CO and TO respectively.  

{\bf LRMI in conventional order} - To determine the CO in a $n+1$D system, we only need to consider regions $A$ and $B$ that are disks $D^n$ with size $r \gg \xi$. The LRMI $\mathcal{I}(D^n)$ is defined in the limit $\text{dist}(A,B)\rightarrow \infty$ with $r$ fixed. Given the ground state Hilbert space $\mathcal{H}_G$ of a gapped system, we propose:
\begin{itemize}
\item[]{\it A gapped system possesses CO when the LRMI $\mathcal{I}(D^n)$ does not vanish for generic states in $\mathcal{H}_G$.}
\end{itemize}
In disordered phases, all correlation functions factorize at large distance, and so does the reduced density matrix $\rho_{A\cup B}$. As such,  LRMI should vanish. Our justification for this proposal is based on the following argument. Consider a basis given by the classical states $\{|i\rangle\}_{i=1}^{\dim (\mathcal{H}_G)}$ with broken symmetries and no connected long-range correlation. For example, in a simple Ising ferromagnet, the ground state Hilbert space is two-dimensional, and the two basis states have all spins up or all spins down. A generic ground state is a superposition $|\psi\rangle = \sum_i \psi_i | i \rangle$. Since the classical states are macroscopically distinguishable, the reduced density matrix of $|\psi\rangle $ on any small region $X$ compared to the system size is block diagonal: $\rho_X= \left(\begin{smallmatrix}|\psi_1|^2 \rho_{1,X} & & \\ & |\psi_2|^2 \rho_{2,X} & \\ & & \ddots\end{smallmatrix}\right)$.
Here $\rho_{i,X}$ is the reduced density matrix of the classical state $|i\rangle$ on the region $X$. Taking $X$ to be the regions $A$, $B$ and $A \cup B$ in the LRMI, we obtain $\mathcal{I} (D^n) = \sum_ i |\psi_1|^2 \mathcal{I}_i (D^n) - \sum_i  |\psi_i|^2 \log  |\psi_i|^2$, where $\mathcal{I}_i (D^n)$ is the LRMI in the classical state $|i\rangle$. Furthermore, $\mathcal{I}_i (D^n) = 0 $ since all the connected correlation functions of the classical state $|i\rangle$ vanish at large distances. Therefore, we obtain
\begin{align}
\mathcal{I} (D^n) = - \sum_i  |\psi_i|^2 \log  |\psi_i|^2,
\label{ShannonE_CO}
\end{align}
which is generically non-vanishing in agreement with our proposal. Moreover, it is worth mentioning that the mutual information $I_{A,B}$ serves as an upper bound for all connected correlation functions\cite{Wolf2008}:
\begin{align}
I_{A,B}\geq \frac{|\langle \mathcal{O}_A \mathcal{O}_B \rangle_c|^2}{2 \|\mathcal{O}_A\|^2 \|\mathcal{O}_B\|^2 },
\label{CorrelUpperBound}
\end{align}
for any operators $\mathcal{O}_A$ and $\mathcal{O}_B$ supported on regions $A$ and $B$ respectively. Here, the subscript ``$c$" stands for connected correlation function and $\|\cdot\|$ for operator 1-norm. Hence, any CO that is signified by the non-vanishing  long-range connected correlation function of the order parameters has a finite value of $\mathcal{I} (D^n)$.

It is worth noting that $\mathcal{I} (D^n)$ in fact takes the form of the Shannon entropy with the probability distribution $\{|\psi_i |^2 \}_{i=1}^{\dim (\mathcal{H}_G)}$. In particular, it can achieve a minimal value of $0$ but only on classical states. This property can be in turn viewed as the definition of the classical states in our LRMI approach, which will later connect to the discussion of the TO. For systems with spontaneous broken continuous symmetries, $\dim (\mathcal{H}_G)$ is infinite. While the non-vanishing LRMI for generic states in $\HG$ is still guaranteed by Eq. \ref{CorrelUpperBound}, its specific form is left for future work.

{\bf LRMI in 2+1D topological order} - In a general $2+1$D system with TO, all the connected correlation functions of local operators vanish at large distance, and so does the LRMI between disks($\mathcal{I} (D^2)$). However, TO on a torus has degenerate ground states, which can be distinguished by non-local processes, e.g., by winding a quasiparticle around the torus. Therefore, it is natural to expect nontrivial LRMI between non-contractible regions, which characterizes string-like order parameters wrapping around the non-contractible loops.
In the following, we focus on the ground state Hilbert space $\HG$ on the torus $T^2$ of size $L_x \times L_y$. The topology of the regions $A$ and $B$ are taken to be $D^1 \times T^1_\gamma$, where $T^1_\gamma$ is an $S^1$ winding around a non-contractible prime 1-cycle $\gamma$ on the $T^2$ (Here a cycle $\gamma$ is defined to be prime when there exists no other cycle such that a multiple $q$ ($q>1$) of it is topologically equivalent to $\gamma$). The linear size $r$ of the $D^1$ part is fixed and satisfies $\xi \ll r \ll L_{x,y}$. The LMRI $\mathcal{I} (D^1 \times T^1_\gamma)$ is defined in the limit where $\distAB \rightarrow \infty$ with $\distAB / L_{x,y}$ fixed. We propose
\begin{itemize}
\item[]{\it A gapped 2+1D system possesses TO (but not CO) when the LRMI $\mathcal{I}(D^1 \times T^1_\gamma)$ is finite for generic states in $\mathcal{H}_G$ with a given non-contractible prime 1-cycle $\gamma$ and $\mathcal{I}(D^2)$ vanishes throughout $\HG$.}
\end{itemize}

\begin{figure}[tb]
\centerline{
\includegraphics[width=3
in]{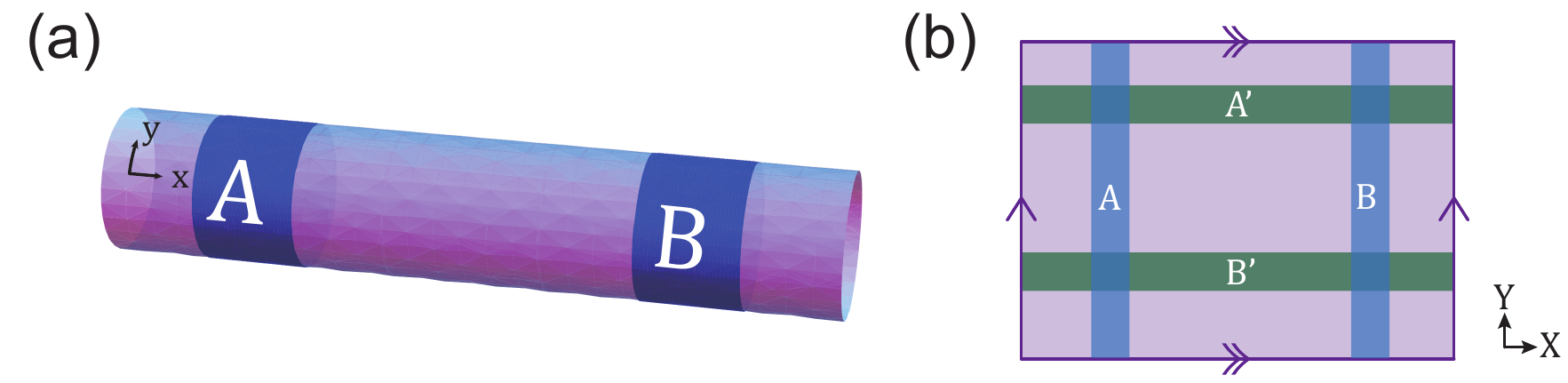}
}
\caption{\label{Ribbons} (a) $\mathcal{I} (D^1 \times T^1_y)$ are the LRMI between regions $A$ and $B$ that wrap around the $y$ direction of the torus. In the thin-torus limit, $\mathcal{I} (D^1)$ reduces to the LRMI $\mathcal{I} (D^1 \times T^1_y)$ that probes the CO in the quasi 1+1D system. (b) On the $T^2$ (edge with the same arrows are identified), the long range limit of the pair $(I_{A,B}, I_{A',B'})$ is constrained by the $\mathcal{S}$-matrix}
\end{figure}

We first consider the thin-torus with $L_y/L_x \ll 1$ limit, in which our proposal can be succinctly explained. The TO on a thin torus can be viewed as the CO of a quasi-(1+1)D system. The most well-known example is the Laughlin fractional quantum Hall state, for which the different topologically degenerate ground states on the $T^2$ cross over to different charge density wave patterns on the thin torus\cite{RezayiHaldane1994,Seidel2005,Seidel2006,Seidel2008,Bergholtz2006,Bergholtz2008}. Another example is given by the 2+1D toric code model which, in the thin-cylinder limit, reduces to a spin chain with $\mathbf{Z}_2 \times \mathbf{Z}_2$ global symmetry. As a quasi 1+1D system, this spin chain exhibits CO with the order parameter given by Wilson loop operators (WLO) along the non-contractible 1-cycle in the $y$ direction \cite{Supplementary}. In the thin-torus limit of a general 2+1D system with TO, the loop order parameters are operators that measure the anyon type in the non-contractible loop along $y$ direction. The long-range correlation of such order parameter simply reflects the fact that the anyon type measured at different $x$ locations should be the same. 
The classical states of this CO are the states with a fixed anyon type, which are known as the minimally entangled states (MES) \cite{Zhang2012MES_supp}. In our LRMI approache, we can choose the LRMI $\mathcal{I} (D^1 \times T^1_y)$ to describe the TO of the 2+1D system (with $\gamma$ the non-contractible 1-cycle along the $y$ direction). In the thin-torus limit (with the ratio $\distAB / L_{y}$ effectively infinity), $\mathcal{I} (D^1 \times T^1_y)$ reduces to the LRMI $\mathcal{I} (D^1)$ that probes the CO in 1+1D system (see Fig. \ref{Ribbons} (a)). Therefore, based on the discussion of the LRMI on system with CO, we conclude that, in the presence of TO, the LRMI $\mathcal{I} (D^1 \times T^1_y)$ in the thin-torus limit stays finite and display the Shannon entropy form for generic states in the $\HG$ and vanishes only on the MES states, which were identified above with the classical states in the quasi-1+1D system.

These results, while initially obtained in the thin-torus imit, remain to hold even when $L_y/L_x \sim 1$. For any non-contractible prime 1-cycle $\gamma$, we can choose the MES basis $\{|i \rangle_\gamma \}_{i=1}^{\dim(\HG)}$ of the $\HG$ with $i$ labeling the anyon types measured by WLOs along the 1-cycle $\gamma$. A generic ground state $|\psi \rangle  \in \HG$ can be expanded as $|\psi \rangle = \sum_i \psi_{\gamma,i} |i \rangle_\gamma$. Similar to the thin-torus limit, the WLO's following the same type of 1-cycle $\gamma$ at different locations exhibit non-trivial correlations which is ensured by Eq. \ref{CorrelUpperBound} to reflect itself in the $\mathcal{I} (D^1 \times T^1_\gamma)$. Using the topological field theory and the replica trick, we can show \cite{Supplementary}
\begin{align}
\mathcal{I} (D^1 \times T^1_\gamma) = - \sum_i  |\psi_{\gamma,i}|^2 \log  |\psi_{\gamma,i}|^2,
\label{ShannonE_TO2}
\end{align}
which is the Shannon entropy form that agrees with our thin-torus analysis and our proposal. Similar to the LRMI redefinition of classical states in CO, Eq. \ref{ShannonE_TO2} allows us to redefine the MES states as those that minimize the LRMI $\mathcal{I} (D^1 \times T^1_\gamma)$.  As we see here, the LRMI unifies the language for the discussion of CO and TO.

In principle, a gapped 2+1D system can possess both CO and TO. While the LRMI $\mathcal{I} (D^2)$ is still a good probe of the CO,  $\mathcal{I} (D^1 \times T^1_\gamma)$ receives contribution from both. The difference of these two are the contribution from TO. 

{\bf Topological uncertainty principle in 2+1D topological order} - We have seen that TO and CO are distinguished by LRMI of different regions. There is another key feature of TO that is different from CO. The loop order parameters in different non-contractible cycles generically do not commute with each other. In other words, a state with a fixed anyon type in one loop is generically a superposition of states with different anyon types in the other loop. For the two prime 1-cycles 
$\gamma_1$ and $\gamma_2$ on the torus, their corresponding MES basis 
$\{|i \rangle_{\gamma_1} \}_{i=1}^{\dim(\HG)}$ and $\{|i \rangle_{\gamma_2} \}_{i=1}^{\dim(\HG)}$ are different and related by the transition matrix $\mathcal{M}(\gamma_1,\gamma_2)_{ij} \equiv _{\gamma_2} \!\!\! \langle j | i \rangle_{\gamma_1}$, also known as the modular matrix that carries the defining data of the TO
\footnote{Non-contractible prime 1-cycles on the $T^2$ can always be mapped to each other by the modular transformation of $T^2$.}.
For example, if we choose $\gamma_{1,2}$ to be the cycles along the $x$ and $y$ direction, $\mathcal{M}(x,y)$ is the modular $\mathcal{S}$-matrix, which is one of the generator of the modular transformation on $T^2$.

\begin{figure}[tb]
\centerline{
\includegraphics[width=3
in]{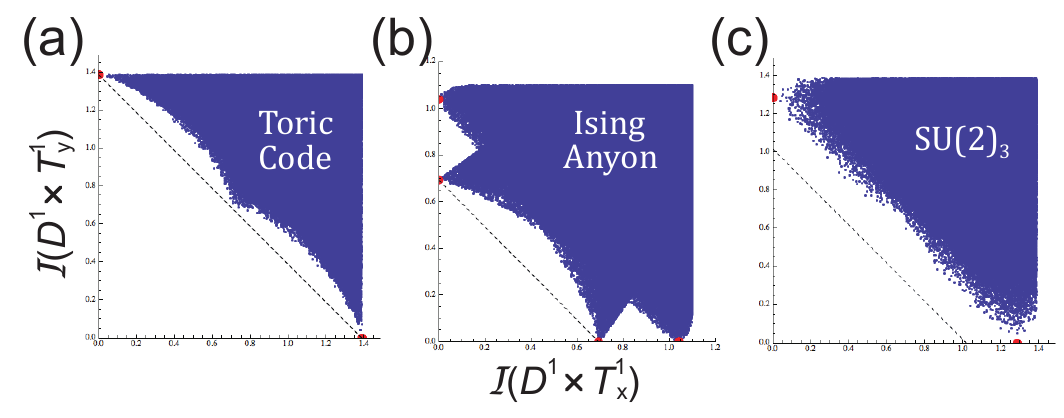}
}
\caption{\label{Crossed_MI} For the toric model (a), the Ising anyon model(b) and the $SU(2)_3$ anyon model (c), the permissible regimes of $(\mathcal{I} (D^1 \times T^1_x),\mathcal{I} (D^1 \times T^1_y))$ are obtained by applying Eq. \ref{ShannonE_TO2} and by Monte-Carlo sampling through $\HG$. }
\end{figure}

The non-trivial modular matrix $\mathcal{M}(\gamma_1,\gamma_2)$ relates the LRMI's $\mathcal{I} (D^1 \times T^1_{\gamma_1})$ and $\mathcal{I} (D^1 \times T^1_{\gamma_2})$ through Eq. \ref{ShannonE_TO2}. In particular, the $\mathcal{S}$-matrix constrains the pair $(\mathcal{I} (D^1 \times T^1_x),\mathcal{I} (D^1 \times T^1_y))$ (which is the long-limit of $(I_{A,B},I_{A',B'})$ shown in Fig. \ref{Ribbons} (b)).
In Fig. \ref{Crossed_MI}, we obtain the permissible pair $(\mathcal{I} (D^1 \times T^1_x),\mathcal{I} (D^1 \times T^1_y))$ by applying Eq. \ref{ShannonE_TO2} and the Monte Carlo sampling through the ground state Hilbert space $\HG$ for well-known theories ((a) 2+1D toric code model, (b) Ising anyon and (c) the $SU(2)_3$ anyon model).
We notice that $\mathcal{I} (D^1 \times T^1_x) \neq \mathcal{I} (D^1 \times T^1_y)$ generically. Furthermore, even though $\mathcal{I} (D^1 \times T^1_x)$ and $\mathcal{I} (D^1 \times T^1_y)$ can individually take values from $0$ to $\log \dim(\HG)$, the sum $\mathcal{I} (D^1 \times T^1_x) + \mathcal{I} (D^1 \times T^1_y)$ is always nonzero, reflecting the fact that no state in $\HG$ can be the MES with respect to both 1-cycles $x$ and $y$. In fact, by applying the Maassen-Uffink inequality\cite{MaassenUffink1988}, we can obtain
\begin{align}
\mathcal{I} (D^1 \times T^1_{\gamma_1}) + \mathcal{I} (D^1 \times T^1_{\gamma_2}) \geq   -2\log \max_{i,j} |
\mathcal{M}{(\gamma_1,\gamma_2)}_{ij}|
\label{TopoUncertainty}
\end{align}
This lower bound is indicated by the dashed line in Fig. \ref{Crossed_MI}. In contrast, the LRMI's in a purely classically ordered system do not depend on the topology of the regions and thus simultaneously vanish for different region choices for the classical states. Therefore Eq .\ref{TopoUncertainty} 
 provides an information theoretic measure of the key difference between TO and CO. A nonzero lower bound of the sum of LRMI in two intersecting pairs of regions measures the 
 non-trivial commutation relations between the WLOs around the two cycles, and guarantees that long-range entanglement is always present in any topological ground state. Therefore, we dub Eq. \ref{TopoUncertainty} the ``{\it topological uncertainty principle}". Although the LRMI and topological uncertainty principle appears to be merely a reformulation of the well-understood theory of CO and (2+1)D TO, it provides a new path for understanding TO's in higher dimensions, as we discuss below.

\begin{figure}[tb]
\centerline{
\includegraphics[width=3
in]{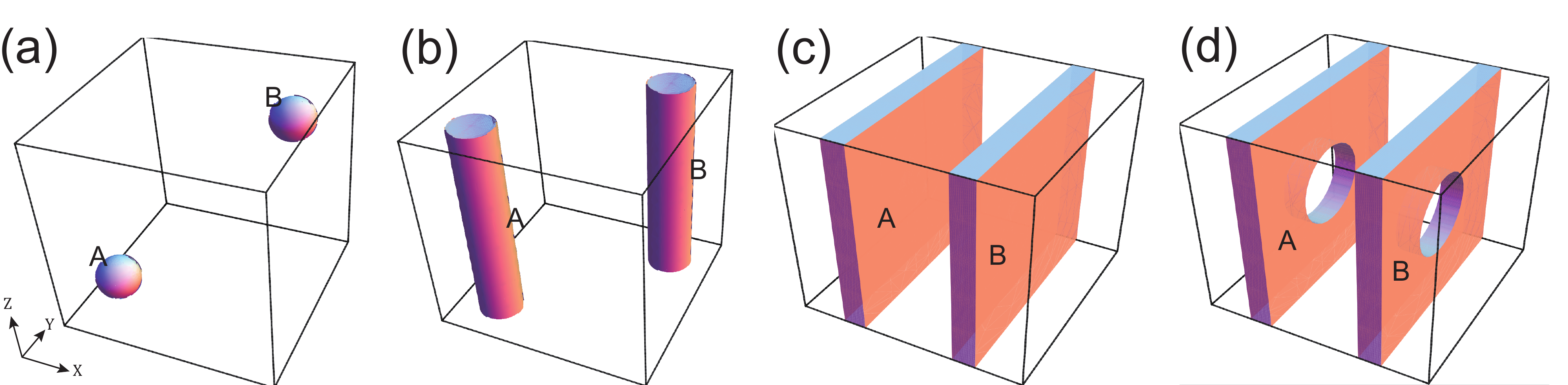}
}
\caption{\label{T3_regions} In 3+1D, the LRMI of different topologies, including (a) $D^3 \approx D^2 \times  (T^1_{\gamma}/D^1)$, (b) $D^2 \times T^1_z$, (c) $D^1 \times T^2_{yz}$ and (d) $D^1 \times (T^2_{yz}/D^2)$, will be considered in the full diagnostic of $m$-membrane condendsate $(m=0,1,2)$. The non-trivial cycles $z$ and $yz$ are chosen here as examples. \label{fig:3D}}
\end{figure}

{\bf LRMI in higher dimensional systems with TO} -
Although TO in higher dimensions have been investigated in some models
\cite{Hastings2005,HammaWen2005,Baez1995,Zeng2015,KongWen2014,Freed1994,WangLevin2015,Levin2005StringNet,Dijkgraaf1990,Joseph2010,LevinStern2011,Bombin2007PRB,Bombin2007PRL,WalkerWang2012,Swingle2011,Keyerlingk2013,WangLevin2014,LinLevin2015,Jian2014,Jiang2014,MoradiWen2015,WangWen2015,Freedman2015}
, the general structure of TO in higher dimensions remains an open question. The understanding of CO and 2+1D TO by LRMI leads to a large class of generalizations--the $m$-membrane condensates in which orders in $n+1$D are induced by condensing $m$ dimensional membranes $(0\leq m<n)$. As we have explained above, CO and 2+1D TO can be respectively viewed as a $0$-membrane (point) condensate and a $1$-membrane (string) condensate. Similarly, $m$-membrane condensates with $m\geq 2$ have been identified in systems with TO in higher dimensions.\cite{Hastings2005,Levin2005StringNet,HammaWen2005,KongWen2014,LinLevin2015,Bombin2007PRB,Bombin2007PRL,Freedman2015} using various of different methods. As we shall see, our LRMI approach offers a unified language for all $m$-membrane condensates in any dimensions.


For $n+1$D systems, we focus on the $n$-torus $T^n$ with a linear size $L_T$, and consider two types of LRMI: $\mathcal{I} (D^{n-m} \times T^m_{\gamma})$ and $\mathcal{I} (D^{n-m} \times (T^m_{\gamma}/D^m))$. Examples for $3+1$D systems are illustrated in Fig. \ref{fig:3D} (c) (d). For $\mathcal{I} (D^{n-m} \times T^m_{\gamma})$, we consider the LRMI between two regions $A$ and $B$ of the same topology $D^{n-m} \times T^m_{\gamma}$ with the $T^m$ part wrapping around a non-contractible prime $m$-cycle $\gamma$ and the linear size $r$ of the $D^{n-m}$ part fixed and satisfying $\xi \ll r \ll L_T$. The
``long range" limit of $\mathcal{I} (D^{n-m} \times T^m_{\gamma})$ is defined as $\distAB \rightarrow \infty$ with $\distAB / L_T$ fixed. $\mathcal{I} (D^{n-m} \times (T^m_{\gamma}/D^m))$ is defined similarly with an extra condition on the (fixed) linear size $r'$ of the $D^m$ part: $\xi \ll r' \ll L_T$. With these definitions, we propose
\begin{itemize}
\item[]{\it A gapped n+1D system possesses an $m$-membrane condensate TO $( 0 \!\leq m \! < n )$ when $\mathcal{I}(D^{n-m} \times T^m_\gamma) > \mathcal{I} (D^{n-m} \times (T^m_{\gamma}/D^m)) \geq 0$ for generic states in $\mathcal{H}_G$ with a given non-contractible prime $m$-cycle $\gamma$.}
\end{itemize}
Here we have adapted the formal notation $D^{n} \times T^0_\gamma \equiv  D^{n}$ and $\mathcal{I} (D^{n} \times (T^0_{\gamma}/D^0)) \equiv 0$.

In the case of $n=2$ and $m=1$, as we notice that $D^{1} \times (T^1_{\gamma}/D^1)$ is topologically equivalent to $D^2$, this proposal is reduced to the proposal for the 2+1D TO. In 3+1D, Fig. \ref{T3_regions} shows examples of regions with different topology relevant to the LRMI: $D^3 \approx D^2 \times  (T^1_{\gamma}/D^1)$, $D^2 \times T^1_z$, $D^1 \times T^2_{yz}$ and $D^1 \times (T^2_{yz}/D^2)$.
Here, the subscripts $z$ and $yz$ represent the non-contractible prime 1-cycle and 2-cycle along the $z$ direction and the $z-y$ plane respectively. As an example,
we can show that the LRMI indicates the coexistence of $1$-membrane and $2$-membrane condensation in the 3+1D toric code model (or equivalently the $\mathbf{Z}_2$ gauge theory) \cite{Supplementary}, which agrees with the result in Ref. \cite{HammaWen2005}.
In 4+1D, two generalized toric code models introduced in Ref. \cite{Dennis2002, Tarun2011} exemplifies two fundamentally different 4+1D TOs: (1) the coexisting 1-membrane and 3-membrane condensates and (2) the 2-membrane condensate \cite{Supplementary}.

In general, an $m$-membrane condensate in $n+1$D should be equipped with the $m$ dimensional Wilson surface operators ($m$-WSO) as its order parameter. The non-trivial long-range correlation function between $m$-WSOs contributes (by Eq. \ref{CorrelUpperBound}) to all LRMI's $\mathcal{I}(D^{n-{m'}} \times T^{m'}_\gamma)$ $(m'\geq m)$ and $D^{n-m''} \times (T^{m''}_\gamma/D^{m''})$ $(m'' > m)$. As we notice that the regions $D^{n-m} \times T^m_\gamma$ and $D^{n-m} \times (T^m_\gamma/D^m)$ contain exactly the same types of WSO (of dimension lower than $m$) but the $m$-WSO (with dimension $m$). Therefore, the difference between $\mathcal{I}(D^{n-m} \times T^m_\gamma)$ and $D^{n-m} \times (T^m_\gamma/D^m)$ only comes from the $m$-membrane condensation.


Similar to the discussion of 2+1D TO, all the TO's signified by non-vanishing $\mathcal{I}(D^{n-m} \times T^m_\gamma)$ can be viewed as the CO in the quasi-$(n-m)+1$D system where the dimensionality associated to the $m$-cycles $\gamma$ are considered as the internal degrees of freedom. Therefore, we generally expect the Shannon entropy form of $\mathcal{I}(D^{n-m} \times T^m_\gamma)$ which is minimized on the eigenbasis of the WSO's contained in the $D^{n-m} \times T^m_\gamma$ region. For two intersecting contractible prime cycles $\gamma_{1,2}$, the WSO's contained in $D^{n-m_1} \times T^{m_1}_{\gamma_1}$ and $D^{n-m_2} \times T^{m_1}_{\gamma_2}$ generically do not commute, resulting in two distinct sets of eigenbasis and, hence, the topological uncertainty principle of $\mathcal{I}(D^{n-m_1} \times T^{m_1}_{\gamma_1}) + \mathcal{I}(D^{n-m_2} \times T^{m_2}_{\gamma_2})$. Here $m_{1,2}$ denotes the dimensionality of the cycles $\gamma_{1,2}$. In particular, this topological uncertainty principle is related to modular transformation on $T^n$ when $m_1=m_2>0$, as it does when $n=2$ and $m_1=m_2=1$.


{\bf Discussion} -
Compared to other entanglement measures of TO in gapped systems, such as the topological entanglement entropy\cite{Kitaev2006EE,Levin2006EE} and other similar universal terms in entanglement entropy\cite{Tarun2011,CasiniF2012,Komargodski2011,Tarun2011N}, the LRMI has the advantage of being ultraviolet finite. All subtleties and ambiguities in entropy calculation due to entanglement of short-scale degrees of freedom\cite{Tarun2011,Liu2013,Casini2014,Casini2015} cancels in LRMI. As the universal terms of the entanglement entropy also appears in gapless systems \cite{Zamolodchikov1986,CasiniF2012,Komargodski2011,FradkinMoore2006,Metlitski2011}, it is interesting to also generalize the LRMI proposal to these cases. In particular, different from disks, the size of the non-contractible regions  have to grow with the system size, resulting possible extra contributions to the LRMI from the gapless mode. The criteria of ordering needs to be modified accordingly.

Ref. \cite{Gaiotto2015} proposed the non-conventional orders induced by the spontaneously broken $m$-form global symmetries in a general dimension. We believe that the LRMI not only captures such non-conventional orders as $m$-membrane condensations but also provides more information because the long-range correlation does not need to be associated with a group structure. As is exemplified by 2+1D TO, $SU(2)_k$ TO states share the same 1-form symmetry\cite{Gaiotto2015} but have distinct LRMI and topological uncertainty relations. It would be important to study the detailed connection between the proposed $m$-membrane condensate and the $m$-form global symmetries.



{\bf Acknowledgement} - We would like to acknowledge the helpful discussion with Horacio Casini. This work was inspired by Xiao-Gang Wen's posting on physics stackexchange \cite{WenStackExchange}.  CMJ and XLQ are supported by the David and Lucile Packard Foundation. XLQ is also supported by the NSF Grant No. DMR-1151786. IK's research at Perimeter Institute is supported in part by the Government of Canada through NSERC and by the Province of Ontario through MRI. We would also like to acknowledge the hospitality of KITP at Santa Barbara, where important progress was made during the program ``Entanglement in Strongly-Correlated Quantum Matter", under NSF Grant No. PHY11-25915.

\bibliography{TI}

\begin{thebibliography}{52}
\expandafter\ifx\csname natexlab\endcsname\relax\def\natexlab#1{#1}\fi
\expandafter\ifx\csname bibnamefont\endcsname\relax
  \def\bibnamefont#1{#1}\fi
\expandafter\ifx\csname bibfnamefont\endcsname\relax
  \def\bibfnamefont#1{#1}\fi
\expandafter\ifx\csname citenamefont\endcsname\relax
  \def\citenamefont#1{#1}\fi
\expandafter\ifx\csname url\endcsname\relax
  \def\url#1{\texttt{#1}}\fi
\expandafter\ifx\csname urlprefix\endcsname\relax\def\urlprefix{URL }\fi
\providecommand{\bibinfo}[2]{#2}
\providecommand{\eprint}[2][]{\url{#2}}

\bibitem[{\citenamefont{Yang}(1962)}]{Yang1962}
\bibinfo{author}{\bibfnamefont{C.~N.} \bibnamefont{Yang}},
  \bibinfo{journal}{Rev. Mod. Phys.} \textbf{\bibinfo{volume}{34}},
  \bibinfo{pages}{694} (\bibinfo{year}{1962}),
  \urlprefix\url{http://link.aps.org/doi/10.1103/RevModPhys.34.694}.

\bibitem[{\citenamefont{Tsui et~al.}(1982)\citenamefont{Tsui, Stormer, and
  Gossard}}]{Tsui1982}
\bibinfo{author}{\bibfnamefont{D.~C.} \bibnamefont{Tsui}},
  \bibinfo{author}{\bibfnamefont{H.~L.} \bibnamefont{Stormer}},
  \bibnamefont{and} \bibinfo{author}{\bibfnamefont{A.~C.}
  \bibnamefont{Gossard}}, \bibinfo{journal}{Phys. Rev. Lett.}
  \textbf{\bibinfo{volume}{48}}, \bibinfo{pages}{1559} (\bibinfo{year}{1982}),
  \urlprefix\url{http://link.aps.org/doi/10.1103/PhysRevLett.48.1559}.

\bibitem[{\citenamefont{Zhang}(1992)}]{Zhang1992}
\bibinfo{author}{\bibfnamefont{S.~C.} \bibnamefont{Zhang}},
  \bibinfo{journal}{International Journal of Modern Physics B}
  \textbf{\bibinfo{volume}{06}}, \bibinfo{pages}{803} (\bibinfo{year}{1992}),
  \urlprefix\url{http://www.worldscientific.com/doi/abs/10.1142/S0217979292000499}.

\bibitem[{\citenamefont{Balents et~al.}(1999)\citenamefont{Balents, Fisher, and
  Nayak}}]{Balents1999}
\bibinfo{author}{\bibfnamefont{L.}~\bibnamefont{Balents}},
  \bibinfo{author}{\bibfnamefont{M.~P.~A.} \bibnamefont{Fisher}},
  \bibnamefont{and} \bibinfo{author}{\bibfnamefont{C.}~\bibnamefont{Nayak}},
  \bibinfo{journal}{Phys. Rev. B} \textbf{\bibinfo{volume}{60}},
  \bibinfo{pages}{1654} (\bibinfo{year}{1999}),
  \urlprefix\url{http://link.aps.org/doi/10.1103/PhysRevB.60.1654}.

\bibitem[{\citenamefont{Levin and Senthil}(2004)}]{LevinSenthil2004}
\bibinfo{author}{\bibfnamefont{M.}~\bibnamefont{Levin}} \bibnamefont{and}
  \bibinfo{author}{\bibfnamefont{T.}~\bibnamefont{Senthil}},
  \bibinfo{journal}{Phys. Rev. B} \textbf{\bibinfo{volume}{70}},
  \bibinfo{pages}{220403} (\bibinfo{year}{2004}),
  \urlprefix\url{http://link.aps.org/doi/10.1103/PhysRevB.70.220403}.

\bibitem[{\citenamefont{Wolf et~al.}(2008)\citenamefont{Wolf, Verstraete,
  Hastings, and Cirac}}]{Wolf2008}
\bibinfo{author}{\bibfnamefont{M.~M.} \bibnamefont{Wolf}},
  \bibinfo{author}{\bibfnamefont{F.}~\bibnamefont{Verstraete}},
  \bibinfo{author}{\bibfnamefont{M.~B.} \bibnamefont{Hastings}},
  \bibnamefont{and} \bibinfo{author}{\bibfnamefont{J.~I.} \bibnamefont{Cirac}},
  \bibinfo{journal}{Phys. Rev. Lett.} \textbf{\bibinfo{volume}{100}},
  \bibinfo{pages}{070502} (\bibinfo{year}{2008}),
  \urlprefix\url{http://link.aps.org/doi/10.1103/PhysRevLett.100.070502}.

\bibitem[{\citenamefont{Rezayi and Haldane}(1994)}]{RezayiHaldane1994}
\bibinfo{author}{\bibfnamefont{E.~H.} \bibnamefont{Rezayi}} \bibnamefont{and}
  \bibinfo{author}{\bibfnamefont{F.~D.~M.} \bibnamefont{Haldane}},
  \bibinfo{journal}{Phys. Rev. B} \textbf{\bibinfo{volume}{50}},
  \bibinfo{pages}{17199} (\bibinfo{year}{1994}),
  \urlprefix\url{http://link.aps.org/doi/10.1103/PhysRevB.50.17199}.

\bibitem[{\citenamefont{Seidel et~al.}(2005)\citenamefont{Seidel, Fu, Lee,
  Leinaas, and Moore}}]{Seidel2005}
\bibinfo{author}{\bibfnamefont{A.}~\bibnamefont{Seidel}},
  \bibinfo{author}{\bibfnamefont{H.}~\bibnamefont{Fu}},
  \bibinfo{author}{\bibfnamefont{D.-H.} \bibnamefont{Lee}},
  \bibinfo{author}{\bibfnamefont{J.~M.} \bibnamefont{Leinaas}},
  \bibnamefont{and} \bibinfo{author}{\bibfnamefont{J.}~\bibnamefont{Moore}},
  \bibinfo{journal}{Phys. Rev. Lett.} \textbf{\bibinfo{volume}{95}},
  \bibinfo{pages}{266405} (\bibinfo{year}{2005}),
  \urlprefix\url{http://link.aps.org/doi/10.1103/PhysRevLett.95.266405}.

\bibitem[{\citenamefont{Seidel and Lee}(2006)}]{Seidel2006}
\bibinfo{author}{\bibfnamefont{A.}~\bibnamefont{Seidel}} \bibnamefont{and}
  \bibinfo{author}{\bibfnamefont{D.-H.} \bibnamefont{Lee}},
  \bibinfo{journal}{Phys. Rev. Lett.} \textbf{\bibinfo{volume}{97}},
  \bibinfo{pages}{056804} (\bibinfo{year}{2006}),
  \urlprefix\url{http://link.aps.org/doi/10.1103/PhysRevLett.97.056804}.

\bibitem[{\citenamefont{Seidel and Yang}(2008)}]{Seidel2008}
\bibinfo{author}{\bibfnamefont{A.}~\bibnamefont{Seidel}} \bibnamefont{and}
  \bibinfo{author}{\bibfnamefont{K.}~\bibnamefont{Yang}},
  \bibinfo{journal}{Phys. Rev. Lett.} \textbf{\bibinfo{volume}{101}},
  \bibinfo{pages}{036804} (\bibinfo{year}{2008}),
  \urlprefix\url{http://link.aps.org/doi/10.1103/PhysRevLett.101.036804}.

\bibitem[{\citenamefont{Bergholtz and Karlhede}(2006)}]{Bergholtz2006}
\bibinfo{author}{\bibfnamefont{E.~J.} \bibnamefont{Bergholtz}}
  \bibnamefont{and} \bibinfo{author}{\bibfnamefont{A.}~\bibnamefont{Karlhede}},
  \bibinfo{journal}{J. Stat. Mech.} p. \bibinfo{pages}{L04001}
  (\bibinfo{year}{2006}),
  \urlprefix\url{http://iopscience.iop.org/1742-5468/2006/04/L04001}.

\bibitem[{\citenamefont{Bergholtz and Karlhede}(2008)}]{Bergholtz2008}
\bibinfo{author}{\bibfnamefont{E.~J.} \bibnamefont{Bergholtz}}
  \bibnamefont{and} \bibinfo{author}{\bibfnamefont{A.}~\bibnamefont{Karlhede}},
  \bibinfo{journal}{Phys. Rev. B} \textbf{\bibinfo{volume}{77}},
  \bibinfo{pages}{155308} (\bibinfo{year}{2008}),
  \urlprefix\url{http://link.aps.org/doi/10.1103/PhysRevB.77.155308}.

\bibitem[{Sup()}]{Supplementary}
\bibinfo{note}{See supplementary material for details}.

\bibitem[{\citenamefont{Maassen and Uffink}(1988)}]{MaassenUffink1988}
\bibinfo{author}{\bibfnamefont{H.}~\bibnamefont{Maassen}} \bibnamefont{and}
  \bibinfo{author}{\bibfnamefont{J.~B.~M.} \bibnamefont{Uffink}},
  \bibinfo{journal}{Phys. Rev. Lett.} \textbf{\bibinfo{volume}{60}},
  \bibinfo{pages}{1103} (\bibinfo{year}{1988}),
  \urlprefix\url{http://link.aps.org/doi/10.1103/PhysRevLett.60.1103}.

\bibitem[{\citenamefont{Hastings and Wen}(2005)}]{Hastings2005}
\bibinfo{author}{\bibfnamefont{M.~B.} \bibnamefont{Hastings}} \bibnamefont{and}
  \bibinfo{author}{\bibfnamefont{X.-G.} \bibnamefont{Wen}},
  \bibinfo{journal}{Phys. Rev. B} \textbf{\bibinfo{volume}{72}},
  \bibinfo{pages}{045141} (\bibinfo{year}{2005}),
  \urlprefix\url{http://link.aps.org/doi/10.1103/PhysRevB.72.045141}.

\bibitem[{\citenamefont{Hamma et~al.}(2005)\citenamefont{Hamma, Zanardi, and
  Wen}}]{HammaWen2005}
\bibinfo{author}{\bibfnamefont{A.}~\bibnamefont{Hamma}},
  \bibinfo{author}{\bibfnamefont{P.}~\bibnamefont{Zanardi}}, \bibnamefont{and}
  \bibinfo{author}{\bibfnamefont{X.-G.} \bibnamefont{Wen}},
  \bibinfo{journal}{Phys. Rev. B} \textbf{\bibinfo{volume}{72}},
  \bibinfo{pages}{035307} (\bibinfo{year}{2005}),
  \urlprefix\url{http://link.aps.org/doi/10.1103/PhysRevB.72.035307}.

\bibitem[{\citenamefont{{Baez} and {Dolan}}(1995)}]{Baez1995}
\bibinfo{author}{\bibfnamefont{J.~C.} \bibnamefont{{Baez}}} \bibnamefont{and}
  \bibinfo{author}{\bibfnamefont{J.}~\bibnamefont{{Dolan}}}, in
  \emph{\bibinfo{booktitle}{eprint arXiv:q-alg/9503002}}
  (\bibinfo{year}{1995}), p. \bibinfo{pages}{3002}.

\bibitem[{\citenamefont{{Zeng} and {Wen}}(2015)}]{Zeng2015}
\bibinfo{author}{\bibfnamefont{B.}~\bibnamefont{{Zeng}}} \bibnamefont{and}
  \bibinfo{author}{\bibfnamefont{X.-G.} \bibnamefont{{Wen}}},
  \bibinfo{journal}{\prb} \textbf{\bibinfo{volume}{91}}, \bibinfo{eid}{125121}
  (\bibinfo{year}{2015}), \eprint{1406.5090}.

\bibitem[{\citenamefont{{Kong} and {Wen}}(2014)}]{KongWen2014}
\bibinfo{author}{\bibfnamefont{L.}~\bibnamefont{{Kong}}} \bibnamefont{and}
  \bibinfo{author}{\bibfnamefont{X.-G.} \bibnamefont{{Wen}}},
  \bibinfo{journal}{ArXiv e-prints}  (\bibinfo{year}{2014}),
  \eprint{1405.5858}.

\bibitem[{\citenamefont{Freed}(1994)}]{Freed1994}
\bibinfo{author}{\bibfnamefont{D.}~\bibnamefont{Freed}},
  \bibinfo{journal}{Communications in Mathematical Physics}
  \textbf{\bibinfo{volume}{159}}, \bibinfo{pages}{343} (\bibinfo{year}{1994}),
  ISSN \bibinfo{issn}{0010-3616},
  \urlprefix\url{http://dx.doi.org/10.1007/BF02102643}.

\bibitem[{\citenamefont{Wang and Levin}(2015)}]{WangLevin2015}
\bibinfo{author}{\bibfnamefont{C.}~\bibnamefont{Wang}} \bibnamefont{and}
  \bibinfo{author}{\bibfnamefont{M.}~\bibnamefont{Levin}},
  \bibinfo{journal}{Phys. Rev. B} \textbf{\bibinfo{volume}{91}},
  \bibinfo{pages}{165119} (\bibinfo{year}{2015}),
  \urlprefix\url{http://link.aps.org/doi/10.1103/PhysRevB.91.165119}.

\bibitem[{\citenamefont{Levin and Wen}(2005)}]{Levin2005StringNet}
\bibinfo{author}{\bibfnamefont{M.~A.} \bibnamefont{Levin}} \bibnamefont{and}
  \bibinfo{author}{\bibfnamefont{X.-G.} \bibnamefont{Wen}},
  \bibinfo{journal}{Phys. Rev. B} \textbf{\bibinfo{volume}{71}},
  \bibinfo{pages}{045110} (\bibinfo{year}{2005}),
  \urlprefix\url{http://link.aps.org/doi/10.1103/PhysRevB.71.045110}.

\bibitem[{\citenamefont{Dijkgraaf and Witten}(1990)}]{Dijkgraaf1990}
\bibinfo{author}{\bibfnamefont{R.}~\bibnamefont{Dijkgraaf}} \bibnamefont{and}
  \bibinfo{author}{\bibfnamefont{E.}~\bibnamefont{Witten}},
  \bibinfo{journal}{Comm. Math. Phys.} \textbf{\bibinfo{volume}{129}},
  \bibinfo{pages}{393} (\bibinfo{year}{1990}),
  \urlprefix\url{http://projecteuclid.org/euclid.cmp/1104180750}.

\bibitem[{\citenamefont{Maciejko et~al.}(2010)\citenamefont{Maciejko, Qi,
  Karch, and Zhang}}]{Joseph2010}
\bibinfo{author}{\bibfnamefont{J.}~\bibnamefont{Maciejko}},
  \bibinfo{author}{\bibfnamefont{X.-L.} \bibnamefont{Qi}},
  \bibinfo{author}{\bibfnamefont{A.}~\bibnamefont{Karch}}, \bibnamefont{and}
  \bibinfo{author}{\bibfnamefont{S.-C.} \bibnamefont{Zhang}},
  \bibinfo{journal}{Phys. Rev. Lett.} \textbf{\bibinfo{volume}{105}},
  \bibinfo{pages}{246809} (\bibinfo{year}{2010}),
  \urlprefix\url{http://link.aps.org/doi/10.1103/PhysRevLett.105.246809}.

\bibitem[{\citenamefont{Levin et~al.}(2011)\citenamefont{Levin, Burnell,
  Koch-Janusz, and Stern}}]{LevinStern2011}
\bibinfo{author}{\bibfnamefont{M.}~\bibnamefont{Levin}},
  \bibinfo{author}{\bibfnamefont{F.~J.} \bibnamefont{Burnell}},
  \bibinfo{author}{\bibfnamefont{M.}~\bibnamefont{Koch-Janusz}},
  \bibnamefont{and} \bibinfo{author}{\bibfnamefont{A.}~\bibnamefont{Stern}},
  \bibinfo{journal}{Phys. Rev. B} \textbf{\bibinfo{volume}{84}},
  \bibinfo{pages}{235145} (\bibinfo{year}{2011}),
  \urlprefix\url{http://link.aps.org/doi/10.1103/PhysRevB.84.235145}.

\bibitem[{\citenamefont{Bombin and
  Martin-Delgado}(2007{\natexlab{a}})}]{Bombin2007PRB}
\bibinfo{author}{\bibfnamefont{H.}~\bibnamefont{Bombin}} \bibnamefont{and}
  \bibinfo{author}{\bibfnamefont{M.~A.} \bibnamefont{Martin-Delgado}},
  \bibinfo{journal}{Phys. Rev. B} \textbf{\bibinfo{volume}{75}},
  \bibinfo{pages}{075103} (\bibinfo{year}{2007}{\natexlab{a}}),
  \urlprefix\url{http://link.aps.org/doi/10.1103/PhysRevB.75.075103}.

\bibitem[{\citenamefont{Bombin and
  Martin-Delgado}(2007{\natexlab{b}})}]{Bombin2007PRL}
\bibinfo{author}{\bibfnamefont{H.}~\bibnamefont{Bombin}} \bibnamefont{and}
  \bibinfo{author}{\bibfnamefont{M.~A.} \bibnamefont{Martin-Delgado}},
  \bibinfo{journal}{Phys. Rev. Lett.} \textbf{\bibinfo{volume}{98}},
  \bibinfo{pages}{160502} (\bibinfo{year}{2007}{\natexlab{b}}),
  \urlprefix\url{http://link.aps.org/doi/10.1103/PhysRevLett.98.160502}.

\bibitem[{\citenamefont{Walker and Wang}(2012)}]{WalkerWang2012}
\bibinfo{author}{\bibfnamefont{K.}~\bibnamefont{Walker}} \bibnamefont{and}
  \bibinfo{author}{\bibfnamefont{Z.}~\bibnamefont{Wang}},
  \bibinfo{journal}{Frontiers of Physics} \textbf{\bibinfo{volume}{7}},
  \bibinfo{pages}{150} (\bibinfo{year}{2012}), ISSN \bibinfo{issn}{2095-0462},
  \urlprefix\url{http://dx.doi.org/10.1007/s11467-011-0194-z}.

\bibitem[{\citenamefont{Swingle et~al.}(2011)\citenamefont{Swingle, Barkeshli,
  McGreevy, and Senthil}}]{Swingle2011}
\bibinfo{author}{\bibfnamefont{B.}~\bibnamefont{Swingle}},
  \bibinfo{author}{\bibfnamefont{M.}~\bibnamefont{Barkeshli}},
  \bibinfo{author}{\bibfnamefont{J.}~\bibnamefont{McGreevy}}, \bibnamefont{and}
  \bibinfo{author}{\bibfnamefont{T.}~\bibnamefont{Senthil}},
  \bibinfo{journal}{Phys. Rev. B} \textbf{\bibinfo{volume}{83}},
  \bibinfo{pages}{195139} (\bibinfo{year}{2011}),
  \urlprefix\url{http://link.aps.org/doi/10.1103/PhysRevB.83.195139}.

\bibitem[{\citenamefont{von Keyserlingk et~al.}(2013)\citenamefont{von
  Keyserlingk, Burnell, and Simon}}]{Keyerlingk2013}
\bibinfo{author}{\bibfnamefont{C.~W.} \bibnamefont{von Keyserlingk}},
  \bibinfo{author}{\bibfnamefont{F.~J.} \bibnamefont{Burnell}},
  \bibnamefont{and} \bibinfo{author}{\bibfnamefont{S.~H.} \bibnamefont{Simon}},
  \bibinfo{journal}{Phys. Rev. B} \textbf{\bibinfo{volume}{87}},
  \bibinfo{pages}{045107} (\bibinfo{year}{2013}),
  \urlprefix\url{http://link.aps.org/doi/10.1103/PhysRevB.87.045107}.

\bibitem[{\citenamefont{Wang and Levin}(2014)}]{WangLevin2014}
\bibinfo{author}{\bibfnamefont{C.}~\bibnamefont{Wang}} \bibnamefont{and}
  \bibinfo{author}{\bibfnamefont{M.}~\bibnamefont{Levin}},
  \bibinfo{journal}{Phys. Rev. Lett.} \textbf{\bibinfo{volume}{113}},
  \bibinfo{pages}{080403} (\bibinfo{year}{2014}),
  \urlprefix\url{http://link.aps.org/doi/10.1103/PhysRevLett.113.080403}.

\bibitem[{\citenamefont{Lin and Levin}(2015)}]{LinLevin2015}
\bibinfo{author}{\bibfnamefont{C.-H.} \bibnamefont{Lin}} \bibnamefont{and}
  \bibinfo{author}{\bibfnamefont{M.}~\bibnamefont{Levin}},
  \bibinfo{journal}{Phys. Rev. B} \textbf{\bibinfo{volume}{92}},
  \bibinfo{pages}{035115} (\bibinfo{year}{2015}),
  \urlprefix\url{http://link.aps.org/doi/10.1103/PhysRevB.92.035115}.

\bibitem[{\citenamefont{Jian and Qi}(2014)}]{Jian2014}
\bibinfo{author}{\bibfnamefont{C.-M.} \bibnamefont{Jian}} \bibnamefont{and}
  \bibinfo{author}{\bibfnamefont{X.-L.} \bibnamefont{Qi}},
  \bibinfo{journal}{Phys. Rev. X} \textbf{\bibinfo{volume}{4}},
  \bibinfo{pages}{041043} (\bibinfo{year}{2014}),
  \urlprefix\url{http://link.aps.org/doi/10.1103/PhysRevX.4.041043}.

\bibitem[{\citenamefont{Jiang et~al.}(2014)\citenamefont{Jiang, Mesaros, and
  Ran}}]{Jiang2014}
\bibinfo{author}{\bibfnamefont{S.}~\bibnamefont{Jiang}},
  \bibinfo{author}{\bibfnamefont{A.}~\bibnamefont{Mesaros}}, \bibnamefont{and}
  \bibinfo{author}{\bibfnamefont{Y.}~\bibnamefont{Ran}},
  \bibinfo{journal}{Phys. Rev. X} \textbf{\bibinfo{volume}{4}},
  \bibinfo{pages}{031048} (\bibinfo{year}{2014}),
  \urlprefix\url{http://link.aps.org/doi/10.1103/PhysRevX.4.031048}.

\bibitem[{\citenamefont{Moradi and Wen}(2015)}]{MoradiWen2015}
\bibinfo{author}{\bibfnamefont{H.}~\bibnamefont{Moradi}} \bibnamefont{and}
  \bibinfo{author}{\bibfnamefont{X.-G.} \bibnamefont{Wen}},
  \bibinfo{journal}{Phys. Rev. B} \textbf{\bibinfo{volume}{91}},
  \bibinfo{pages}{075114} (\bibinfo{year}{2015}),
  \urlprefix\url{http://link.aps.org/doi/10.1103/PhysRevB.91.075114}.

\bibitem[{\citenamefont{Wang and Wen}(2015)}]{WangWen2015}
\bibinfo{author}{\bibfnamefont{J.~C.} \bibnamefont{Wang}} \bibnamefont{and}
  \bibinfo{author}{\bibfnamefont{X.-G.} \bibnamefont{Wen}},
  \bibinfo{journal}{Phys. Rev. B} \textbf{\bibinfo{volume}{91}},
  \bibinfo{pages}{035134} (\bibinfo{year}{2015}),
  \urlprefix\url{http://link.aps.org/doi/10.1103/PhysRevB.91.035134}.

\bibitem[{\citenamefont{{Freedman} and {Hastings}}(2015)}]{Freedman2015}
\bibinfo{author}{\bibfnamefont{M.~H.} \bibnamefont{{Freedman}}}
  \bibnamefont{and} \bibinfo{author}{\bibfnamefont{M.~B.}
  \bibnamefont{{Hastings}}}, \bibinfo{journal}{ArXiv e-prints}
  (\bibinfo{year}{2015}), \eprint{1507.05676}.

\bibitem[{\citenamefont{Dennis et~al.}(2002)\citenamefont{Dennis, Kitaev,
  Landahl, and Preskill}}]{Dennis2002}
\bibinfo{author}{\bibfnamefont{E.}~\bibnamefont{Dennis}},
  \bibinfo{author}{\bibfnamefont{A.}~\bibnamefont{Kitaev}},
  \bibinfo{author}{\bibfnamefont{A.}~\bibnamefont{Landahl}}, \bibnamefont{and}
  \bibinfo{author}{\bibfnamefont{J.}~\bibnamefont{Preskill}},
  \bibinfo{journal}{Journal of Mathematical Physics}
  \textbf{\bibinfo{volume}{43}} (\bibinfo{year}{2002}).

\bibitem[{\citenamefont{Grover et~al.}(2011)\citenamefont{Grover, Turner, and
  Vishwanath}}]{Tarun2011}
\bibinfo{author}{\bibfnamefont{T.}~\bibnamefont{Grover}},
  \bibinfo{author}{\bibfnamefont{A.~M.} \bibnamefont{Turner}},
  \bibnamefont{and}
  \bibinfo{author}{\bibfnamefont{A.}~\bibnamefont{Vishwanath}},
  \bibinfo{journal}{Phys. Rev. B} \textbf{\bibinfo{volume}{84}},
  \bibinfo{pages}{195120} (\bibinfo{year}{2011}),
  \urlprefix\url{http://link.aps.org/doi/10.1103/PhysRevB.84.195120}.

\bibitem[{\citenamefont{Kitaev and Preskill}(2006)}]{Kitaev2006EE}
\bibinfo{author}{\bibfnamefont{A.}~\bibnamefont{Kitaev}} \bibnamefont{and}
  \bibinfo{author}{\bibfnamefont{J.}~\bibnamefont{Preskill}},
  \bibinfo{journal}{Phys. Rev. Lett.} \textbf{\bibinfo{volume}{96}},
  \bibinfo{pages}{110404} (\bibinfo{year}{2006}),
  \urlprefix\url{http://link.aps.org/doi/10.1103/PhysRevLett.96.110404}.

\bibitem[{\citenamefont{Levin and Wen}(2006)}]{Levin2006EE}
\bibinfo{author}{\bibfnamefont{M.}~\bibnamefont{Levin}} \bibnamefont{and}
  \bibinfo{author}{\bibfnamefont{X.-G.} \bibnamefont{Wen}},
  \bibinfo{journal}{Phys. Rev. Lett.} \textbf{\bibinfo{volume}{96}},
  \bibinfo{pages}{110405} (\bibinfo{year}{2006}),
  \urlprefix\url{http://link.aps.org/doi/10.1103/PhysRevLett.96.110405}.

\bibitem[{\citenamefont{Casini and Huerta}(2012)}]{CasiniF2012}
\bibinfo{author}{\bibfnamefont{H.}~\bibnamefont{Casini}} \bibnamefont{and}
  \bibinfo{author}{\bibfnamefont{M.}~\bibnamefont{Huerta}},
  \bibinfo{journal}{Phys. Rev. D} \textbf{\bibinfo{volume}{85}},
  \bibinfo{pages}{125016} (\bibinfo{year}{2012}),
  \urlprefix\url{http://link.aps.org/doi/10.1103/PhysRevD.85.125016}.

\bibitem[{\citenamefont{Komargodski and Schwimmer}(2011)}]{Komargodski2011}
\bibinfo{author}{\bibfnamefont{Z.}~\bibnamefont{Komargodski}} \bibnamefont{and}
  \bibinfo{author}{\bibfnamefont{A.}~\bibnamefont{Schwimmer}},
  \bibinfo{journal}{Journal of High Energy Physics}
  \textbf{\bibinfo{volume}{2011}}, \bibinfo{eid}{99} (\bibinfo{year}{2011}),
  \urlprefix\url{http://dx.doi.org/10.1007/JHEP12%282011%29099}.

\bibitem[{\citenamefont{{Grover}}(2011)}]{Tarun2011N}
\bibinfo{author}{\bibfnamefont{T.}~\bibnamefont{{Grover}}},
  \bibinfo{journal}{ArXiv e-prints}  (\bibinfo{year}{2011}),
  \eprint{1112.2215}.

\bibitem[{\citenamefont{Liu and Mezei}(2013)}]{Liu2013}
\bibinfo{author}{\bibfnamefont{H.}~\bibnamefont{Liu}} \bibnamefont{and}
  \bibinfo{author}{\bibfnamefont{M.}~\bibnamefont{Mezei}},
  \bibinfo{journal}{Journal of High Energy Physics}
  \textbf{\bibinfo{volume}{2013}}, \bibinfo{eid}{162} (\bibinfo{year}{2013}),
  \urlprefix\url{http://dx.doi.org/10.1007/JHEP04%282013%29162}.

\bibitem[{\citenamefont{Casini et~al.}(2014)\citenamefont{Casini, Huerta, and
  Rosabal}}]{Casini2014}
\bibinfo{author}{\bibfnamefont{H.}~\bibnamefont{Casini}},
  \bibinfo{author}{\bibfnamefont{M.}~\bibnamefont{Huerta}}, \bibnamefont{and}
  \bibinfo{author}{\bibfnamefont{J.~A.} \bibnamefont{Rosabal}},
  \bibinfo{journal}{Phys. Rev. D} \textbf{\bibinfo{volume}{89}},
  \bibinfo{pages}{085012} (\bibinfo{year}{2014}),
  \urlprefix\url{http://link.aps.org/doi/10.1103/PhysRevD.89.085012}.

\bibitem[{\citenamefont{{Casini} et~al.}(2015)\citenamefont{{Casini}, {Huerta},
  {Myers}, and {Yale}}}]{Casini2015}
\bibinfo{author}{\bibfnamefont{H.}~\bibnamefont{{Casini}}},
  \bibinfo{author}{\bibfnamefont{M.}~\bibnamefont{{Huerta}}},
  \bibinfo{author}{\bibfnamefont{R.~C.} \bibnamefont{{Myers}}},
  \bibnamefont{and} \bibinfo{author}{\bibfnamefont{A.}~\bibnamefont{{Yale}}},
  \bibinfo{journal}{ArXiv e-prints}  (\bibinfo{year}{2015}),
  \eprint{1506.06195}.

\bibitem[{\citenamefont{Zamolodchikov}(1986)}]{Zamolodchikov1986}
\bibinfo{author}{\bibfnamefont{A.~B.} \bibnamefont{Zamolodchikov}},
  \bibinfo{journal}{JETP Lett.} \textbf{\bibinfo{volume}{43}},
  \bibinfo{pages}{730} (\bibinfo{year}{1986}), \bibinfo{note}{[Pisma Zh. Eksp.
  Teor. Fiz.43,565(1986)]}.

\bibitem[{\citenamefont{Fradkin and Moore}(2006)}]{FradkinMoore2006}
\bibinfo{author}{\bibfnamefont{E.}~\bibnamefont{Fradkin}} \bibnamefont{and}
  \bibinfo{author}{\bibfnamefont{J.~E.} \bibnamefont{Moore}},
  \bibinfo{journal}{Phys. Rev. Lett.} \textbf{\bibinfo{volume}{97}},
  \bibinfo{pages}{050404} (\bibinfo{year}{2006}),
  \urlprefix\url{http://link.aps.org/doi/10.1103/PhysRevLett.97.050404}.

\bibitem[{\citenamefont{{Metlitski} and {Grover}}(2011)}]{Metlitski2011}
\bibinfo{author}{\bibfnamefont{M.~A.} \bibnamefont{{Metlitski}}}
  \bibnamefont{and} \bibinfo{author}{\bibfnamefont{T.}~\bibnamefont{{Grover}}},
  \bibinfo{journal}{ArXiv e-prints}  (\bibinfo{year}{2011}),
  \eprint{1112.5166}.

\bibitem[{\citenamefont{Gaiotto et~al.}(2015)\citenamefont{Gaiotto, Kapustin,
  Seiberg, and Willett}}]{Gaiotto2015}
\bibinfo{author}{\bibfnamefont{D.}~\bibnamefont{Gaiotto}},
  \bibinfo{author}{\bibfnamefont{A.}~\bibnamefont{Kapustin}},
  \bibinfo{author}{\bibfnamefont{N.}~\bibnamefont{Seiberg}}, \bibnamefont{and}
  \bibinfo{author}{\bibfnamefont{B.}~\bibnamefont{Willett}},
  \bibinfo{journal}{Journal of High Energy Physics}
  \textbf{\bibinfo{volume}{2015}}, \bibinfo{eid}{172} (\bibinfo{year}{2015}),
  \urlprefix\url{http://dx.doi.org/10.1007/JHEP02%282015%29172}.

\bibitem[{\citenamefont{Wen}()}]{WenStackExchange}
\bibinfo{author}{\bibfnamefont{X.-G.} \bibnamefont{Wen}}, \bibinfo{note}{{W}hat
  is spontaneous symmetry breaking in QUANTUM systems?, {URL}
  http://physics.stackexchange.com/questions/29311/what-is-spontaneous-symmetry-breaking-in-quantum-systems}.

\end{thebibliography}


\begin{thebibliography}{8}
\expandafter\ifx\csname natexlab\endcsname\relax\def\natexlab#1{#1}\fi
\expandafter\ifx\csname bibnamefont\endcsname\relax
  \def\bibnamefont#1{#1}\fi
\expandafter\ifx\csname bibfnamefont\endcsname\relax
  \def\bibfnamefont#1{#1}\fi
\expandafter\ifx\csname citenamefont\endcsname\relax
  \def\citenamefont#1{#1}\fi
\expandafter\ifx\csname url\endcsname\relax
  \def\url#1{\texttt{#1}}\fi
\expandafter\ifx\csname urlprefix\endcsname\relax\def\urlprefix{URL }\fi
\providecommand{\bibinfo}[2]{#2}
\providecommand{\eprint}[2][]{\url{#2}}

\bibitem[{\citenamefont{Zhang et~al.}(2012)\citenamefont{Zhang, Grover, Turner,
  Oshikawa, and Vishwanath}}]{Zhang2012MES_supp}
\bibinfo{author}{\bibfnamefont{Y.}~\bibnamefont{Zhang}},
  \bibinfo{author}{\bibfnamefont{T.}~\bibnamefont{Grover}},
  \bibinfo{author}{\bibfnamefont{A.}~\bibnamefont{Turner}},
  \bibinfo{author}{\bibfnamefont{M.}~\bibnamefont{Oshikawa}}, \bibnamefont{and}
  \bibinfo{author}{\bibfnamefont{A.}~\bibnamefont{Vishwanath}},
  \bibinfo{journal}{Phys. Rev. B} \textbf{\bibinfo{volume}{85}},
  \bibinfo{pages}{235151} (\bibinfo{year}{2012}),
  \urlprefix\url{http://link.aps.org/doi/10.1103/PhysRevB.85.235151}.

\bibitem[{\citenamefont{Dong et~al.}(2008)\citenamefont{Dong, Fradkin, Leigh,
  and Nowling}}]{Dong2008_supp}
\bibinfo{author}{\bibfnamefont{S.}~\bibnamefont{Dong}},
  \bibinfo{author}{\bibfnamefont{E.}~\bibnamefont{Fradkin}},
  \bibinfo{author}{\bibfnamefont{R.~G.} \bibnamefont{Leigh}}, \bibnamefont{and}
  \bibinfo{author}{\bibfnamefont{S.}~\bibnamefont{Nowling}},
  \bibinfo{journal}{Journal of High Energy Physics}
  \textbf{\bibinfo{volume}{2008}}, \bibinfo{pages}{016} (\bibinfo{year}{2008}),
  \urlprefix\url{http://stacks.iop.org/1126-6708/2008/i=05/a=016}.

\bibitem[{\citenamefont{Witten}(1989)}]{Witten1989_supp}
\bibinfo{author}{\bibfnamefont{E.}~\bibnamefont{Witten}},
  \bibinfo{journal}{Comm. Math. Phys.} \textbf{\bibinfo{volume}{121}},
  \bibinfo{pages}{351} (\bibinfo{year}{1989}),
  \urlprefix\url{http://projecteuclid.org/euclid.cmp/1104178138}.

\bibitem[{\citenamefont{Kitaev and Preskill}(2006)}]{Kitaev2006EE_supp}
\bibinfo{author}{\bibfnamefont{A.}~\bibnamefont{Kitaev}} \bibnamefont{and}
  \bibinfo{author}{\bibfnamefont{J.}~\bibnamefont{Preskill}},
  \bibinfo{journal}{Phys. Rev. Lett.} \textbf{\bibinfo{volume}{96}},
  \bibinfo{pages}{110404} (\bibinfo{year}{2006}),
  \urlprefix\url{http://link.aps.org/doi/10.1103/PhysRevLett.96.110404}.

\bibitem[{\citenamefont{Levin and Wen}(2006)}]{Levin2006EE_supp}
\bibinfo{author}{\bibfnamefont{M.}~\bibnamefont{Levin}} \bibnamefont{and}
  \bibinfo{author}{\bibfnamefont{X.-G.} \bibnamefont{Wen}},
  \bibinfo{journal}{Phys. Rev. Lett.} \textbf{\bibinfo{volume}{96}},
  \bibinfo{pages}{110405} (\bibinfo{year}{2006}),
  \urlprefix\url{http://link.aps.org/doi/10.1103/PhysRevLett.96.110405}.

\bibitem[{\citenamefont{Hamma et~al.}(2005)\citenamefont{Hamma, Zanardi, and
  Wen}}]{HammaWen2005_supp}
\bibinfo{author}{\bibfnamefont{A.}~\bibnamefont{Hamma}},
  \bibinfo{author}{\bibfnamefont{P.}~\bibnamefont{Zanardi}}, \bibnamefont{and}
  \bibinfo{author}{\bibfnamefont{X.-G.} \bibnamefont{Wen}},
  \bibinfo{journal}{Phys. Rev. B} \textbf{\bibinfo{volume}{72}},
  \bibinfo{pages}{035307} (\bibinfo{year}{2005}),
  \urlprefix\url{http://link.aps.org/doi/10.1103/PhysRevB.72.035307}.

\bibitem[{\citenamefont{Dennis et~al.}(2002)\citenamefont{Dennis, Kitaev,
  Landahl, and Preskill}}]{Dennis2002_supp}
\bibinfo{author}{\bibfnamefont{E.}~\bibnamefont{Dennis}},
  \bibinfo{author}{\bibfnamefont{A.}~\bibnamefont{Kitaev}},
  \bibinfo{author}{\bibfnamefont{A.}~\bibnamefont{Landahl}}, \bibnamefont{and}
  \bibinfo{author}{\bibfnamefont{J.}~\bibnamefont{Preskill}},
  \bibinfo{journal}{Journal of Mathematical Physics}
  \textbf{\bibinfo{volume}{43}} (\bibinfo{year}{2002}).

\bibitem[{\citenamefont{Grover et~al.}(2011)\citenamefont{Grover, Turner, and
  Vishwanath}}]{Tarun2011_supp}
\bibinfo{author}{\bibfnamefont{T.}~\bibnamefont{Grover}},
  \bibinfo{author}{\bibfnamefont{A.~M.} \bibnamefont{Turner}},
  \bibnamefont{and}
  \bibinfo{author}{\bibfnamefont{A.}~\bibnamefont{Vishwanath}},
  \bibinfo{journal}{Phys. Rev. B} \textbf{\bibinfo{volume}{84}},
  \bibinfo{pages}{195120} (\bibinfo{year}{2011}),
  \urlprefix\url{http://link.aps.org/doi/10.1103/PhysRevB.84.195120}.

\end{thebibliography}

\def\ShannonTO{4}

\onecolumngrid

\newpage
\vspace{1cm}
\begin{center}
{\bf\Large Supplementary material}
\end{center}
\vspace{0.2cm}

\setcounter{equation}{0}
\setcounter{figure}{0}

\renewcommand{\thetable}{S\Roman{table}}
\renewcommand{\thefigure}{S\arabic{figure}}
\renewcommand{\thesection}{S\Roman{section}}
\renewcommand{\thesubsection}{S\arabic{subsection}}
\renewcommand{\theequation}{S\arabic{equation}}

This supplementary material contains 4 sections. In the first section, we review the definition of the 2+1D toric code model and study its thin-torus limit, demonstrating the connection between the 2+1D topological order and the conventional order in 1+1D. The second section provides a derivation of the long-range mutual information (LRMI) Eq. $\ShannonTO$ for the 2-torus and generalize this equation to a general 2-manifold. In the third section, we study the LRMI in the 3+1D toric code model. We show that this model exhibits both 1-membrane and 2-membrane condensations. In last section, we study the LRMI in 2 different 4+1D toric code models with one of them hosting a coexisting 1-membrane and 3-membrane condensations while the other exhibits the condensation of 2-membranes.

\section{Thin-torus limit of the 2+1D toric code model}
\begin{figure}[tb]
\centerline{
\includegraphics[width=5
in]{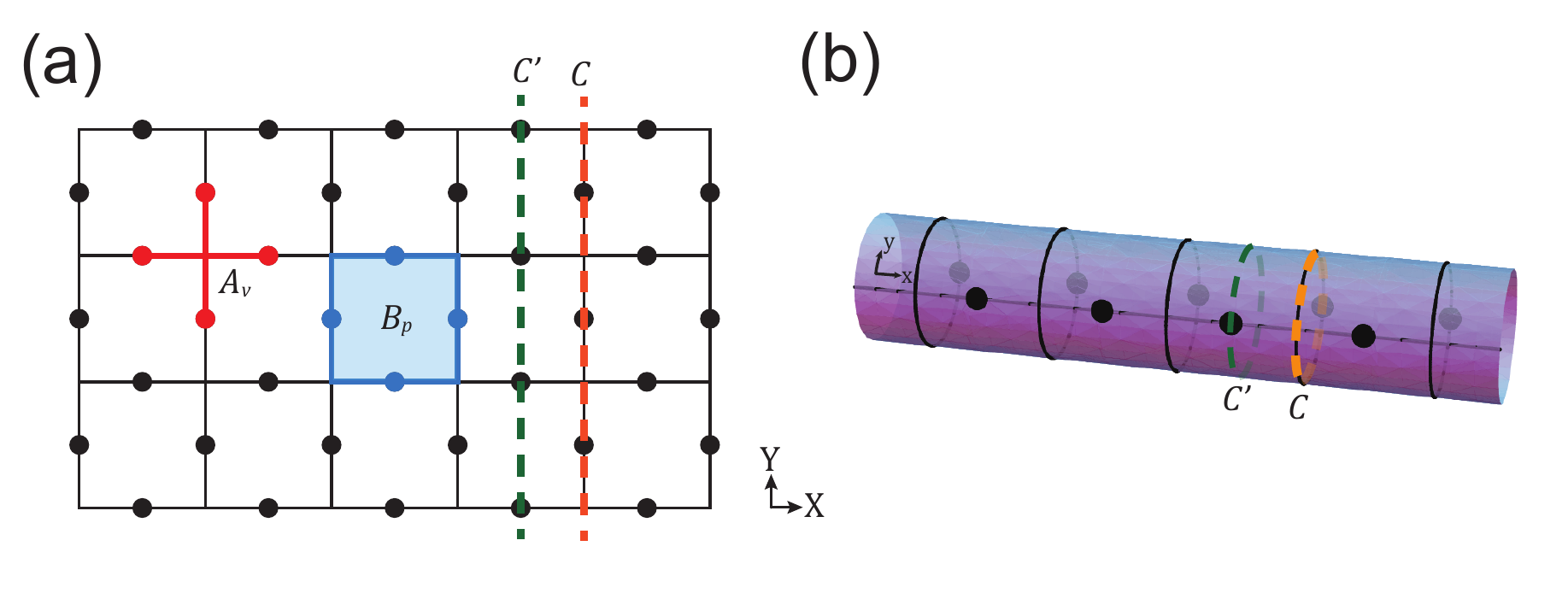}
}
\caption{\label{TC} (a) The 2+1D toric code model is defined on a 2D square lattice with the degrees of freedom (black dots) on the links. In its Hamiltonian $H_\text{TC}$, the vertex term $A_v$ acts on the 4 links (red) connected to the vertex $v$, while the plaquette term $B_p$ acts on the 4 links (blue) on the boundary of the plaquette $p$. The Wilson loop operator $W_e(C)$ and $W_m(C')$ are defined as a product of spin operators along the path $C$ (orange) and $C'$ (green). (b) In the thin-torus limit where $L_y=1$ in the lattice unit, the 2+1D toric code model is reduced to a spin chain shown in the figure. The WLO (orange and green), $W_e(C)$ and $W_m(C')$,  reduce to single spin operators and the local order parameter of the spin chain.}
\end{figure}

In this section, we will show how the $\mathbf{Z}_2$ topological order (TO) of the 2+1D toric code model reduces to the conventional order (CO) of the quasi 1+1D system in the thin-torus limit. The 2+1D toric code model is defined on a 2D square lattice with the degrees of freedom on the links and the Hamiltonian given by
\begin{align}
H_\text{TC} = - \sum_v A_v - \sum_p B_p,
\end{align}
where $v$ and $p$ labels the vertices and the plaquettes respectively (see Fig. \ref{TC} (a)). The vertex term $A_v$ and the plaquette term $B_p$ are given by
\begin{align}
A_v =\prod_{l \in v} \sigma_l^x,~~~  B_p =\prod_{l \in p} \sigma_l^z
\end{align}
where the index $l$ labels the links that connect to the vertex $v$ or belong to the plaquette $p$ and $\sigma^{x,z}$ denote the Pauli matrices. All the vertex terms $A_v$ and plaquette terms $B_p$ commute with each other, making this model exactly solvable. A ground state $|\psi \rangle $ of $H_\text{TC}$ should satisfy
\begin{align}
A_v|\psi\rangle=B_p|\psi\rangle=|\psi\rangle,
\end{align}
for any vertex $v$ and plaquette $p$. The ground state of $H_\text{TC}$ on an $S^2$ is non-degenerate. However, there are 4 degenerate ground states on the torus $T^2$. This 4-fold ground state degeneracy is a signature of the TO. The 4 degenerate ground state can not be distinguished by local operators, while the non-contractible Wilson loop operators (WLO) act non-trivially among them. The WLO are given by
\begin{align}
W_e(C)= \prod_{l\in C} \sigma_i^z,~~~ W_m(C')= \prod_{l\in C'} \sigma_i^x,
\end{align}
where $C$ and $C'$ are both non-contractible closed path on the $T^2$, one along the links of the lattice and the other along the links of the dual lattice (see Fig. \ref{TC} (a)). It is easy to show that
\begin{align}
[W_e(C), H_\text{TC} ] = 0,~~~ [W_m(C'), H_\text{TC} ] = 0.
\end{align}
When acting on the ground state Hilbert space $\HG$, these WLO's $W_e(C)$ and $W_m(C')$ do not depend on the detail of the path $C$ and $C'$, but rather only on the type of non-contractible 1-cycles of $T^2$ the path $C$ and $C'$ traverse along. In Fig. \ref{TC} (a), we have chosen the path $C$ and $C'$ to follow the non-contractible 1-cycle along the $y$ direction. In this case, they commute with each other and each of them have eigenvalues $\pm 1$. The 4 simultaneous eigen states of $W_e(C)$ and $W_m(C')$ with 4 different combination of the eigenvalues serve as a basis of $\HG$ on the torus $T^2$. These eigenvalues in fact label the different anyon types in the TO threading through the non-contractible 1-cycle along the $y$ direction. Therefore, as is proposed in Ref. \cite{Zhang2012MES_supp}, this basis coincide with the basis of the minimally entangled states (MES).

Now, we consider the thin-torus limit of the $2+1$D toric code model. Because the $2+1$D toric code Hamiltonian consists of only commuting terms, the correlation length of the system is in fact $0$. Therefore, we can, without loss of the topological nature of the system, take the thin-torus limit with the circumference in the $y$ direction $L_y=1$ in the lattice unit (see Fig. \ref{TC} (b)). In this limit, the model is reduced to two decoupled Ising chains with the Hamiltonian
\begin{align}
H_\text{1dTC} = - \sum_{i\in \mathbf{Z}} \sigma^x_i   \sigma^x_{i+1} - \sum_{i\in \mathbf{Z}} \sigma^z_{i-1/2}   \sigma^z_{i+1/2}.
\label{Ham1dTC}
\end{align}
The subscripts $i$ and $i+1/2$ with $i\in \mathbf{Z}$ label the links along the $x$ and $y$ direction respectively. Obviously, the system has a global $\mathbf{Z}_2 \times \mathbf{Z}_2$ symmetry and the ground states possess the conventional order as 1+1D system. In this limit, the Wilson loop operators $W_e(C)$ and $W_m(C')$ are reduced to single-spin operators $\sigma^x_i$ and $\sigma^z_{i+1/2}$, which are exactly the order parameters of the 1+1D system. Therefore, the MES of the 2+1D TO are identified with the classical states in the 1+1D conventional order.

\section{Long-range mutual information for 2+1D topologically ordered systems on the 2-torus $T^2$ and other general 2-manifolds}
In the main text, we present Eq. $\ShannonTO$ for the long-range mutual information (LRMI) in 2+1D topologically ordered system on a 2-torus $T^2$. In this section, we provide a field theoretic derivation of Eq. $\ShannonTO$ and generalize it to a general 2-manifold. First, we review the calculation of the entanglement entropy for 2+1D topological states in a general setting by using the replica trick \cite{Dong2008_supp}. Consider a general 2+1D topological system with the degrees of freedom given by the field $\varphi(x)$ and the (Euclidean) field theory action given by $\S_\text{E}[\varphi(x)]$. In the limit where the physical correlation length $\xi \rightarrow 0$, the (unnormalized) ground state wave functional $\Psi[\varphi(x)]$ on a closed 2-manifold $\mathcal{N}$ can be written as a path integral on any 3-manifold $\mathcal{B}$ such that $\partial\mathcal{B}=\mathcal{N}$ with $\varphi(x)$ as the boundary condition:
\begin{align}
\Psi[\varphi(x)] =  \int_{y \in \mathcal{B}, \varphi'(y) \big|_{\partial B} = \varphi(y) } \text{D}[\varphi'(y)] e^{-\S_\text{E}[\varphi'(y)]} .
\label{Wave_Funtional}
\end{align}
The (unnormalized) reduced density matrix of a region $\mathcal{X}$ on the 2-manifold $\mathcal{N}$ can also be expressed in the path integral form. Consider two copies of $\mathcal{B}$ denoted as $\mathcal{B}^1$ and $\mathcal{B}^2$. The corresponding regions on their boundaries are denoted as $\mathcal{X}^1$ and $\mathcal{X}^2$. We partially glue together the 3-manifolds $\mathcal{B}^1$ and $\mathcal{B}^2$ along $\partial \mathcal{B}^1/ \mathcal{X}^1 $ and $\partial \mathcal{B}^2/ \mathcal{X}^2 $, creating a new 3-manifold $\mathcal{R}$ with its boundary $\partial \mathcal{R}= \mathcal{X}^1 \cup \mathcal{X}^2$. The (unnormalized) reduced density matrix can be written as
\begin{align}
\rho_\mathcal{X}[\varphi(x), \varphi'(x)] = \Tr_{\mathcal{N}/\mathcal{X}} |\Psi \rangle\langle \Psi| = \int_{y \in \mathcal{R}, ~ \varphi''(y) \big|_{ \mathcal{X}^1} = \varphi(y), ~ \varphi''(y) \big|_{ \mathcal{X}^2} = \varphi'(y) }  \text{D}[\varphi''(y)] e^{-\S_\text{E}[\varphi''(y)]} .
\label{ReducedDM_FT}
\end{align}
The entanglement entropy of the region $\mathcal{X}$ can be obtained as
\begin{align}
S_\mathcal{X} = -\lim_{n\rightarrow 1}\frac{\text{d}}{ \text{d}n } \left( \frac{\text{Tr} \rho_\mathcal{X}^n}{\left(\text{Tr} \rho_\mathcal{X}\right)^n} \right)
\label{Renyi_Limit}
\end{align}
We can express $\Tr \rho_\mathcal{N}^n$ as a partition function $\mathcal{Z}(\mathcal{R}_n)$ on the replica manifold $\mathcal{R}_n$. $\mathcal{R}_n$ is obtained from gluing $n$ copies of the 3-manifold $\mathcal{R}$ where the $\mathcal{X}^2$ part of the boundary $\partial\mathcal{R}$ of the $i$th copy is identified with the $\mathcal{X}^1$ part of the boundary $\partial\mathcal{R}$ of the $i+1$th copy (with $i=1,2,...,n$ and the identification $n+1 \sim 1$). Therefore, we obtain the entanglement entropy as
\begin{align}
S_\mathcal{X} = -\lim_{n\rightarrow 1}\frac{\text{d}}{ \text{d}n } \left( \frac{\mathcal{Z}(\mathcal{R}_n)}{\mathcal{Z}(\mathcal{R}_1)^n} \right).
\end{align}

\subsection{Derivation of the LRMI Eq. $\ShannonTO$ on 2-torus $T^2$}
\begin{figure}[tb]
\centerline{
\includegraphics[width=6
in]{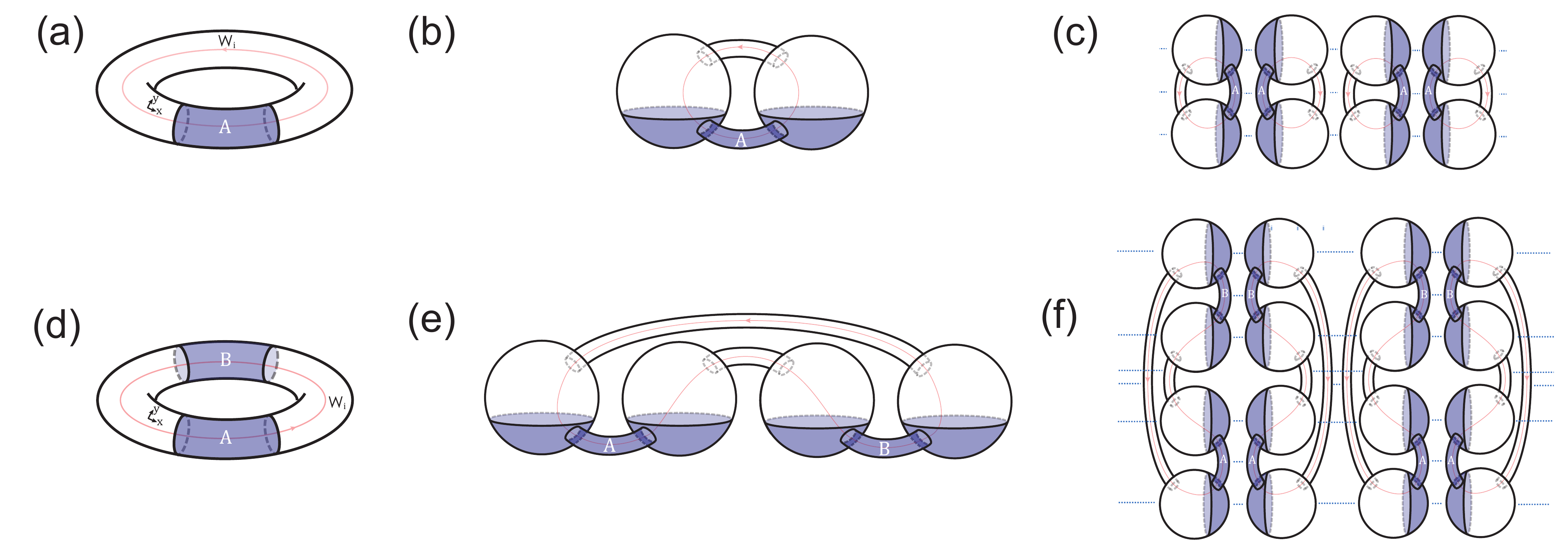}
}
\caption{\label{Replica} (a) The MES $|i\rangle_y$ on the $T^2$ with the $i$th type threading threding through the non-contractible 1-cycle along the $y$ direction can be generated by the path integral on the solid torus $\mathcal{B}=D^2\times T^1$ with the WLO $\mathsf{W}_i$ inserted. The $D^1 \times T^1_y$ region $A$ of the $T^2$ is depicted. (b) shows a 3-manifold that is topologically
equivalent to the solid torus $\mathcal{B}=D^2\times T^1$. (c) shows the identification (indicated by the blue dotted lines) in the process of gluing $2n$ copies of $\mathcal{B}$ into the replica manifold $\mathcal{R}_n$. (d) depicts the two separated regions $A$ and $B$ on the $T^2$, each with the topology of $D^1 \times T^1_y$. (e) shows a 3 manifold that is topologically equivalent to the one shown in (d). (f) shows the identification (indicated by the blue dotted lines) in the process of gluing $2n$ copies of $\mathcal{B}$ into the replica manifold $\tilde{\mathcal{R}}_n$.
 }
\end{figure}

Eq. $\ShannonTO$ applies to the LRMI $\mathcal{I}_\psi (D^1 \times T^1_\gamma)$ for any non-contractible prime 1-cycle $\gamma$ on $T^2$. Since any non-contractible prime 1-cycle $\gamma$ can be mapped to each other by the modular transformations of the $T^2$, it is sufficient to derive Eq. $\ShannonTO$ just for the LRMI $\mathcal{I}_\psi (D^1 \times T^1_\gamma)$ for a specific 1-cycle $\gamma$ which is chosen here to be the non-contractible prime 1-cycle along the $y$ direction. For this purpose, we need to calculate the entanglement entropy $S_A$, $S_B$ and $S_{A \cup B}$ for regions $A$ and $B$ shown in Fig. \ref{Replica} (d). Ref. \cite{Dong2008_supp} has already presented the derivation of $S_A$ and $S_B$ which we will review for the sake of completeness. Then, we generalize this calculation to
$S_{A \cup B}$ and derive the LRMI $\mathcal{I}_\psi (D^1 \times T^1_y)$.

Now we restrict our discussion to the ground states on the 2-torus $\mathcal{N}=T^2$ with the corresponding 3-manifold $\mathcal{B}=D^2 \times S^1$ a solid torus. We will assume that a 2+1D topological system always admit a Chern-Simons theory description. This assumption allows us to construct different degenerate ground state on the $T^2$ by inserting different WLO's into the path integral on $\mathcal{B}$\cite{Witten1989_supp} (see Fig. \ref{Replica} (a)):
\begin{align}
\Psi_i[\varphi(x)] =  \int_{y \in \mathcal{B}, \varphi'(y) \big|_{\partial B} = \varphi(x) } \text{D}[\varphi'(y)] \mathsf{W}_i e^{-\S_\text{E}[\varphi'(y)]}.
\label{Wave_Funtional_T2}
\end{align}
This wave functional $\Psi_i[\varphi(x)]$ exactly represents the MES state $|i\rangle_y$ (defined in the main text) with $i$th type anyon threading through the non-contractible 1-cycle along the $y$ direction.

\begin{figure}[tb]
\centerline{
\includegraphics[width=4
in]{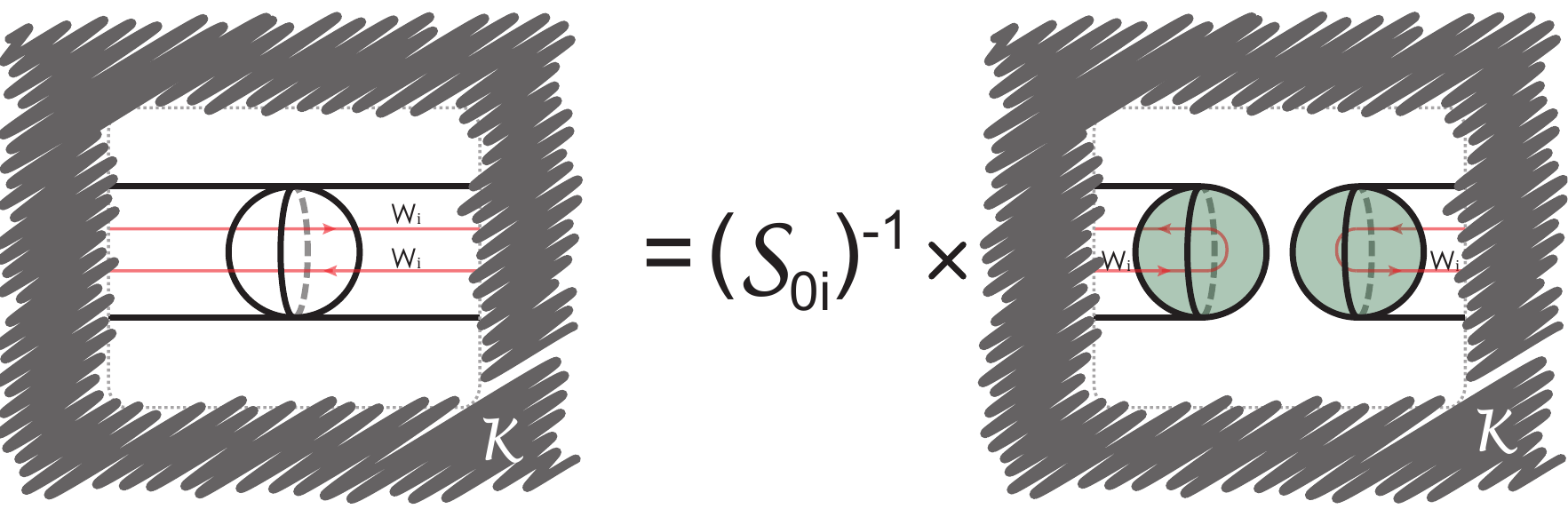}
}
\caption{\label{surgery} The surgery procedure relates, by a factor of $\mathcal{S}^{-1}_{0j}$, the partitions on two different closed 3-manifolds. The 3-manifold on the left has $\mathbb{R} \times S^2$ structure (a ``tube" structure) in a certain neighborhood and the rest of the 3-manifold is denoted as $\mathcal{K}$. The 3-manifold on the right is obtained from the previous one by switching the ``tube" structure to $((\mathbb{R}^- \times S^2) / (\{0^-\} \times S^2)) \cup ((\mathbb{R}^+ \times S^2) / (\{ 0^+ \} \times S^2))$ and reconnecting the WLO's without changing $\mathcal{K}$. This procedure effectively disconnects the ``tube" structure into two pieces with the two open ends capped off by two 3-disks $D^3$.
}
\end{figure}

Following Ref. \cite{Dong2008_supp}, we will calculate the entanglement entropy $S_A$ of the $D^1\times T^1_y$ region $A$ on $T^2$ (see Fig. \ref{Replica} (a)). We first focus on the entanglement entropy $S_{i,A}$ on the MES state $|i\rangle_y$ (or equivalently $\Psi_i[\varphi(x)]$). The solid torus $\mathcal{B}=D^2 \times S^1$ from which $\Psi_i[\varphi(x)]$ is obtained is topologically equivalent to 2 3-disks $D^3$ connected by two ``bridges" $S^2 \times D^1$ (see Fig. \ref{Replica} (b)). We denote the reduced density matrix of $|i\rangle_y$ on the region $A$ is denoted as $\rho_{i,A}$. The quantity $\Tr (\rho_{i,A})^n$ in Eq. \ref{Renyi_Limit} is equivalent to the partition function on the replica manifold $\mathcal{R}_n$ obtained from gluing $2n$ copies of the solid torus $\mathcal{B}= D^2 \times S^1$ together. The gluing process is done by identifying the boundary of these $2n$ solid tori according to the dotted blue lines shown in Fig. \ref{Replica} (c). In Fig. \ref{Replica} (c), the 3-disks on the upper row are glued together into an 3-sphere $S^3$ with $2n$ 3-disks $D^3$ punctures. The same applies to the 3-disks on the lower row. The $4n$ ``bridges" are pairwise glued into $2n$ ``tubes" of the topology $S^2 \times D^1$, connecting the two 3-spheres $S^3$ through their $D^3$ punctures. Therefore, the replica manifold $\mathcal{R}_n$ is a manifold with 2 3-spheres $S^3$ connected by $2n$ ``tubes". The partition function on $\mathcal{R}_n$ can be evaluated through surgery \cite{Dong2008_supp,Witten1989_supp}. As is shown in Fig. \ref{surgery}, each of the ``tubes" that connects the two 3-spheres $S^3$ can be disconnected into two pieces with the two open ends capped off by two 3-disks $D^3$ and the Wilson loops reconnected accordingly. The partition function of the manifolds before and after this surgery process differ by a factor of $(\mathcal{S}_{0i})^{-1}$, where $\mathcal{S}$ is the topological $\mathcal{S}$-matrix, $i$ is the anyon type the WLO $\mathsf{W}_i$ carries and $0$ represents the trivial anyon. After disconnecting all the ``tubes" in $\mathcal{R}_n$, we obtain two disconnected $S^3$ each carrying a WLO $\mathsf{W}_i$, the partition function of which is given by $|\mathcal{S}_{0i}|^2$. Therefore, we have
\begin{align}
\Tr (\rho_{i,A}^n) = |\mathcal{S}_{0i}|^{2(1-n)}.
\end{align}
Applying Eq. \ref{Renyi_Limit}, we obtain
\begin{align}
S_{i,A}=-\lim_{n\rightarrow 1}\frac{\text{d}}{ \text{d}n } \left( \frac{\text{Tr} \rho_{i,A}^n}{\left(\text{Tr} \rho_{i,A}\right)^n} \right) = 2 \log |\mathcal{S}_{0i}|,
\label{MES_EE}
\end{align}
which is the topological part of the entanglement entropy \cite{Kitaev2006EE_supp, Levin2006EE_supp}. The usual ``area law" term in the entanglement entropy vanishes in this calculation because the application of the Chern-Simons theory effectively projects out the higher energy physics, leading to a $\xi=0$ correlation length.

Now we can caculate the entanglement entropy $S_{\psi,A}$ for a generic state $|\psi\rangle$ in the ground state Hilbert space $\HG$ on $T^2$. We can expand the state $|\psi\rangle$ in the MES basis: $|\psi \rangle = \sum_i \psi_{y,i} |i\rangle_y$. Its reduced density matrix $\rho_{\psi, A}$ on the region $A$ then takes a block diagonal form: $\left(
\begin{array}{ccc}
|\psi_1|^2 \rho_{1,A} & & \\
 & |\psi_2|^2 \rho_{2,A} & \\
 & & \ddots
\end{array}
\right)$. Therefore, the entanglement entropy is given by
\begin{align}
S_{\psi,A} & = \sum_i  |\psi_{y,i}|^2 S_{i,A} - \sum_i  |\psi_{y,i}|^2 \log |\psi_{y,i}|^2 \nonumber \\
& = \sum_i 2 |\psi_{y,i}|^2 \log |\mathcal{S}_{0i}|  - \sum_i  |\psi_{y,i}|^2 \log |\psi_{y,i}|^2.
\end{align}
With the same calculation, we can show that the entanglement entropy $S_{\psi,B}$ for the region $B$ is identical to $S_{\psi,A}$.

In the following, we will generalize the calculation of $S_{\psi,A}$ studied in Ref. \cite{Dong2008_supp} to the entanglement entropy $S_{\psi,A \cup B}$ for the regions $A \cup B$ shown in Fig. \ref{Replica} (d). Again, we first focus on the MES state $|i\rangle_y$ ( or $\Psi_i[\varphi(x)]$). The solid torus $\mathcal{B}=D^2 \times S^1$ from which the wave functional $\Psi_i[\varphi(x)]$ is obtained is topologically equivalent to 4 3-disks $D^3$ connected by four ``bridges" $S^2 \times D^1$ (see Fig. \ref{Replica} (e)). The reduced density matrix of $|i\rangle_y$ on the region $A \cup B$ is denoted as $\rho_{i,A \cup B}$. $\Tr (\rho_{i,A\cup B}^n)$ can be identified as the partition function on the replica manifold $\tilde{\mathcal{R}}_n$ obtained from gluing $2n$ copies of the solid torus $\mathcal{B}=D^2 \times S^1$ together in the way indicated in Fig. \ref{Replica} (f). Similar to the discussion on $\mathcal{R}_n$, we can view the replica manifold $\tilde{\mathcal{R}}_n$ as 4 3-spheres $S^3$, each with $2n$ 3-disks $D^3$ punctures,  connected in a cyclic order by in total $4n$ ``tubes" $S^2 \times D^1$. Through surgery, we can obtain the partition function on $\tilde{\mathcal{R}}_n$ and therefore $\Tr (\rho_{i,A \cup B} ^n )$ as
\begin{align}
\Tr (\rho_{i,A \cup B} ^n ) = \tilde{\mathcal{R}}_n =|\mathcal{S}_{0i}|^{4(1-n)}.
\end{align}
Tthe entanglement entropy is then given by
\begin{align}
S_{i,A \cup B} = 4 \log |\mathcal{S}_{0i}|.
\end{align}
For a generic state $|\psi \rangle = \sum_i \psi_{y,i} |i\rangle_y$ in the ground state Hilbert space, the reduced density matrix $\rho_{\psi, A\cup B}$ on the region $A \cup B$, similar to $\rho_{\psi, A}$, also takes a block diagonal form. Therefore, the entanglement entropy is given by
\begin{align}
S_{\psi,A \cup B} & = \sum_i  |\psi_{y,i}|^2 S_{i,A\cup B} - \sum_i  |\psi_{y,i}|^2 \log |\psi_{y,i}|^2 \nonumber \\
& = \sum_i 4 |\psi_{y,i}|^2 \log |\mathcal{S}_{0i}|  - \sum_i  |\psi_{y,i}|^2 \log |\psi_{y,i}|^2.
\end{align}
Since we have taken the limit of vanishing correlation length, the LRMI $\mathcal{I}_\psi (D^1 \times T^1_\gamma)$ is directly given by the mutual information between regions $A$ and $B$ :
\begin{align}
\mathcal{I}_\psi (D^1 \times T^1_y) =  S_{\psi,A } + S_{\psi, B} -S_{\psi,A \cup B} = -\sum_i |\psi_{y,i}|^2 \log |\psi_{y,i}|^2,
\label{ShannonE}
\end{align}
which is exactly the Eq. $\ShannonTO$ of the main text. In the presence of a finite correlation length $\xi$, the mutual information between $A$ and $B$ has an extra contribution from the local correlations with expected form $ \sim L_y e^{-\distAB / \xi} $. In the definition of LRMI, we consider the limit with $\distAB / L_{x,y}$ finite and $\distAB \rightarrow \infty$. The contribution from the local correlations vanishes in this limit. Therefore, the result of $\mathcal{I}_\psi (D^1 \times T^1_y) $ is still valid. It is worth mentioning that Eq. \ref{ShannonE} allow us alternatively define the MES basis $\{|i\rangle_y\}$ as the states that minimizes the LRMI $\mathcal{I}_\psi (D^1 \times T^1_y)$.

\subsection{Derivation of the LRMI $\mathcal{I}_\psi (D^1 \times T^1_\gamma)$ on a general 2-manifold}
\begin{figure}[tb]
\centerline{
\includegraphics[width=5
in]{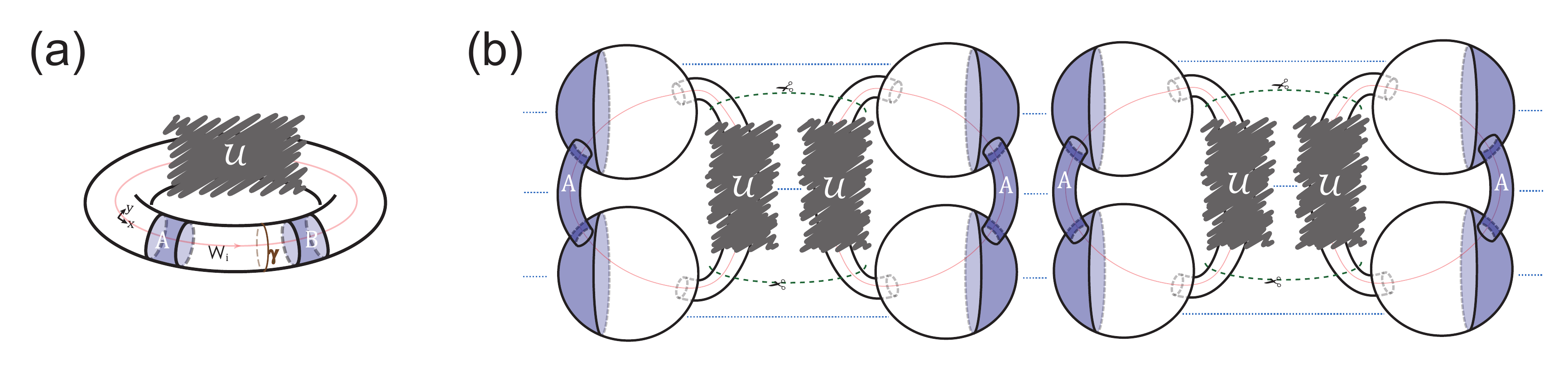}
}
\caption{\label{Replica_MultiT} (a) shows the manifold $\mathcal{N}$ with a neighborhood containing the region $A$ and $B$ and with the rest of manifold denoted as $\mathcal{U}$. (b) shows the identification (indicated by the blue dotted lines) in the process of gluing $2n$ copies of $\mathcal{B}$ into the replica manifold $\mathcal{R}_n$. The surgery procedure along the green dotted lines allows us to isolate the contribution from $\mathcal{U}$ and its interior to the partition function.}
\end{figure}

In fact, the derivation of the LRMI $\mathcal{I}_\psi (D^1 \times T^1_\gamma)$ in the previous subsection can be generalized to a general 2-manifolds $\mathcal{N}$ with any non-contractible prime 1-cycles $\gamma$. We can always deform the 2-manifolds $\mathcal{N}$ such that the neighborhood containing the two disconnected $D^1 \times T^1_\gamma$ region $A$ and $B$ coincides with Fig. \ref{ReducedDM_FT} (a). Here $\mathcal{U}$ represents the rest of the manifold $\mathcal{N}$ outside of this neighborhood. Let us first consider a state $|i,u_i\rangle$ with a fixed anyon type $i$ threading through the 1-cycle $\gamma$. The wave function on $\mathcal{U}$
is specified by the label $u_i$. The path integral on $\mathcal{B}$ that generates this state $|i,u_i\rangle$ should contain a WLO $\mathsf{W}_i$ as is shown in Fig. \ref{Replica_MultiT} (a). The entanglement entropy $S_{i,A}$ of the state $|i,u_i\rangle$ on the region $A$ can be calculated through $\Tr (\rho_{i,A})^n $. $\Tr (\rho_{i,A}^n) $ which is equivalent to the partition function on the replica manifold $\mathcal{R}_n$ obtained from gluing $2n$ copies of $\mathcal{B}$ together in the way indicated by Fig. \ref{Replica_MultiT} (b). Notice that, by performing the surgery procedure along the green dotted line in Fig. \ref{Replica_MultiT} (b), we can isolate the contribution from the $2n$ copies of the $\mathcal{U}$ part (including its interior) of the replica manifold $\mathcal{R}_n$ and obtain
\begin{align}
\Tr (\rho_{i,A}^n) = |\mathcal{S}_{0i}|^{2-n} \mathcal{F}_\mathcal{U}(|i,u_i\rangle)^{2n},
\end{align}
where $\mathcal{F}_\mathcal{U}(|i,u_i\rangle)$ is the contribution from one copy of the $\mathcal{U}$ (and its interior), which depends on the quantum state $|i,u_i\rangle$. Regardless of the specific form of $\mathcal{F}_\mathcal{U}$, we always have
\begin{align}
\frac{\Tr (\rho_{i,A}^n)}{\Tr (\rho_{i,A})^n} = |\mathcal{S}_{0i}|^{2(1-n)},
\end{align}
and, thus, the entanglement entropy is given by
\begin{align}
S_{i,A}=2 \log |\mathcal{S}_{0i}|,
\end{align}
which depends only on the anyon type $i$ but not the other wave functions label $u_i$. We also recognize that this formula is exactly the same as result Eq. \ref{MES_EE} on $T^2$. The same result applies to the suregion $B$: $S_{i,B}=2 \log |\mathcal{S}_{0i}|$. A similar derivation can be done to obtain the entanglement entropy $S_{i,A\cup B}$ of the state $|i,u_i\rangle$ on the region $A \cup B$:
\begin{align}
S_{i,A\cup B}=4 \log |\mathcal{S}_{0i}|,
\end{align}
which again coincides with the result on the $T^2$ and only depends on the anyon type $i$.

A generic state $|\psi\rangle$ on $\mathcal{N}$ can always be expanded in the basis with fixed anyon type threading through the non-contractible 1-cycle $\gamma$: $|\psi\rangle=\sum_i \psi_{\gamma,i}|i,u_i\rangle $. Similar to the case of $\mathcal{N}=T^2$, all three reduced density matrix $\rho_{\psi,A}$, $\rho_{\psi,B}$ and $\rho_{\psi,A \cup B}$ are block diagonal (with each block labeled by the anyon type), following which we have
\begin{align}
\mathcal{I}_\psi (D^1 \times T^1_\gamma) & = \sum_i  |\psi_{\gamma,i}|^2 (S_{i,A}+S_{i,B}-S_{i,A\cup B}) - \sum_i  |\psi_{\gamma,i}|^2 \log |\psi_{\gamma,i}|^2
\nonumber \\
 & = -\sum_i |\psi_{\gamma,i}|^2 \log |\psi_{\gamma,i}|^2.
\end{align}
This result again follows the Shannon entropy (SE) form on the probability distribution $\{|\psi_{\gamma,i}|^2\}_i$ of the quantum state $|\psi\rangle$ on the anyon type threading through the non-contractible 1-cycle $\gamma$. The LRMI $\mathcal{I}_\psi (D^1 \times T^1_y)$ is still minimized on the MES state on a general 2-manifold $\mathcal{N}$

\section{Long-range mutual information in the 3+1D toric code model}

\begin{figure}[tb]
\centerline{
\includegraphics[width=4
in]{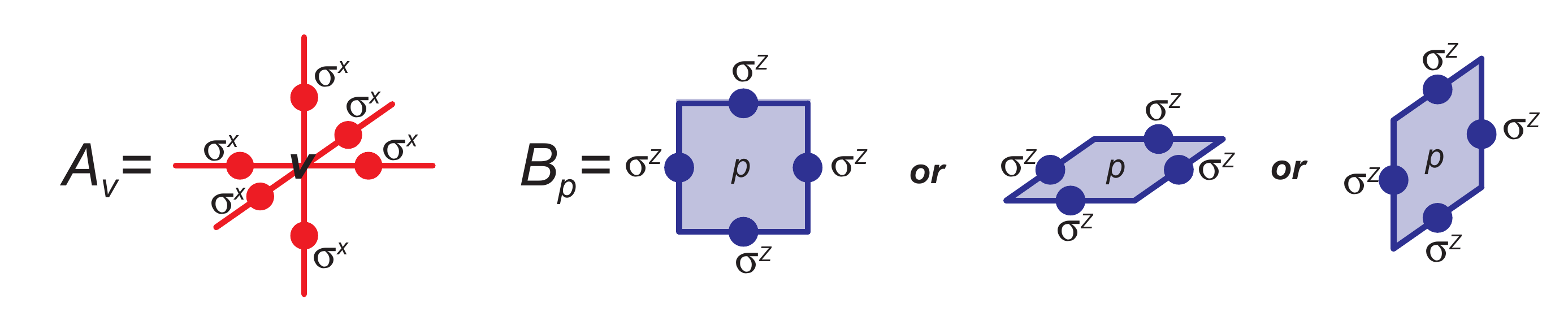}
}
\caption{\label{Toric3D_Ham} The vertex term $A_v$ and the plquette terms $B_p$ are depiced in this figure. }
\end{figure}

In this section, we will study the long-range mutual information of the 3+1D toric code model on the 3-torus $T^3$. The 3+1D toric code model is defined on a cubic lattice with the degrees of freedom on the links and the Hamiltonian given by
\begin{align}
H_\text{3DTC} = - \sum_v A_v - \sum_p B_p,
\end{align}
where $v$ and $p$ labels the vertices and the plaquettes. The vertex term $A_v$ and the plaquette term $B_p$ in the Hamiltonian $H_\text{3DTC}$ are given by (see in Fig. \ref{Toric3D_Ham})
\begin{align}
A_v =\prod_{l \in v} \sigma_l^x, ~~  B_p =\prod_{l \in p} \sigma_l^z,
\end{align}
where the index $l$ labels the links that connect to the vertex $v$ or belong to the plaquette $p$. All terms in the Hamiltonian commute with each other rendering this model exactly solvable. Any ground state $|\psi \rangle$ of the Hamiltonian should satisfy
\begin{align}
A_v|\psi\rangle=B_p|\psi\rangle=|\psi\rangle,
\end{align}
for any vertex $v$ and any plaquette $p$. The ground state is non-degenerate on a 3-sphere. However, the ground state degeneracy on 3-torus is $8$-fold indicating the existence of TO. Within the ground state Hilbert space $\HG$ of the 3 torus $T^3$, the WLO $W_{x,y,z}$ and 2-WSO's $W_{xy,yz,zx}$ acts non-trivially. (Here we've adopted the notation ``2-WSO" from the main text.) $W_x$ is a product of $\sigma^z$ along the links which form a path that winds around the non-contractible prime 1-cycle along the $x$ direction. $W_{y,z}$ are defined similarly. $W_{xy}$ is a product of $\sigma^x$ on the links whose dual plaquettes form a 2D surface wrapping around the non-contractible prime 2-cycle along the $x-y$ plane. $W_{yz,zx}$ are defined similarly.

In the following, we will follow the proposal in the main text and show that the 3+1D toric model exhibits 1-membrane and 2-membrane condensation by computing the following LRMI on the 3-torus $T^3$: $\mathcal{I}(D^3)$, $\mathcal{I}(D^2 \times T^1_z)$, $\mathcal{I}(D^1 \times T^2_{yz})$ and $\mathcal{I}(D^1 \times (T^2_{yz}/D^2))$. These calculations are done for a generic ground state $|\psi\rangle$ in the ground state Hilbert space $\HG$ on the $T^3$. It is worth mentioning that Ref. \cite{HammaWen2005} also pointed out that the 3+1D toric model can be viewed as a condensate of ``strings" and ``membrane" (which correspond to 1-membrane and 2-membrane in our proposal). However, our definition of condensate is completely different from, and should be more general than the study presented in Ref.   \cite{HammaWen2005_supp}. Before getting into the detailed calculations, we first set up the convention for the choice of regions and terminology. As a convention, we define a region to be a collection of cubic cells such that if a cubic cell belongs to this region, so do all of its vertices, links and plaquettes. Also, if a vertex $v$, a link $l$ or a plaquette $p$ belongs to the region, there must exist a cube in this region such that this cube contains $v$, $l$ or $p$ respectively. A vertex is defined as a boundary vertex of a region if the vertex $v$ belongs to the region, but there exist links outside of this regions that are connected to $v$.

\subsection{Calculation of the LRMI $\mathcal{I}(D^3)$}
We consider two separated $D^3$ regions $A$ and $B$. The entanglement entropy $S_{\psi,A}$,  $S_{\psi,B}$ and $S_{\psi,A \cup B}$ associated to the regions $A$, $B$ and $A\cup B$ can be obtained by the Schmidt decomposition of $|\psi\rangle$. First, we consider the region $A$. For each boundary vertex $v \in \partial A$, we define the boundary vertex operator $\tilde{A}_v = \prod_{l \in v ~\&~ l\in A } \sigma_l^x$, a product of $\sigma^x$'s on the links that both belong to the region $A$ and are connected to the vertex $v$. A boundary condition on $\partial A$ is defined by the set $\{\tilde{a}_v\}_{v\in \partial A}$, which is a choice of eigenvalues $\tilde{a}_v\pm 1$ for all the boundary vertex operators $\tilde{A}_v$. The ground state $|\psi\rangle$ admits a Schmidt decomposition:
\begin{align}
|\psi\rangle= \sum_i \lambda_{\{\tilde{a}_v\}} |\psi^A_{\{\tilde{a}_v\}} \rangle |\psi^{\bar{A}}_{\{\tilde{a}_v\}} \rangle,
\end{align}
with the Schmidt state $|\psi^A_{\{\tilde{a}_v\}} \rangle$ defined on the region $A$ satisfying the boundary conditions
\begin{align}
\tilde{A}_v |\psi^A_{\{\tilde{a}_v\}} \rangle = \tilde{a}_v |\psi^A_{\{\tilde{a}_v\}}\rangle,~~~ \forall v\in \partial A,
\end{align}
and the ``ground state conditions"
\begin{align}
A_v |\psi^A_{\{\tilde{a}_v\}} \rangle &=  |\psi^A_{\{\tilde{a}_v\}}\rangle, ~~~ \forall v\in (A/\partial A)
\nonumber \\
B_p |\psi^A_{\{\tilde{a}_v\}} \rangle &=  |\psi^A_{\{\tilde{a}_v\}}\rangle, ~~~ \forall p\in A.
\end{align}
Since the number of equations above and the number of degrees of freedom contained in the region $A$, the Schmidt state $|\psi^A_{\{\tilde{a}_v\}} \rangle$ is uniquely determined by these equations, which confirms that the one-to-one correspondence between the Schmidt states and the choices of boundary conditions. Notice the identity $\prod_{v\in \partial A} \tilde{A}_v = \prod _{v\in (A/\partial A)} A_v$. A boundary condition is permissible only when
\begin{align}
\prod_{v\in \partial A} \tilde{a}_v=1.
\end{align}
The number of permissible boundary conditions is then given by $2^{|\partial A|-1}$ with $|\partial A|$ the number of boundary vertices.
Also notice that, for the plaquettes $p$ that share links with but don't belong to the region $A$, the action of $B_p$ leaves the ground state $|\psi\rangle$ invariant, but maps the Schmidt state with one boundary condition to another. The Schmidt coefficients $\lambda_{\{\tilde{a}_v\}}$ of boundary conditions that are related by these plaquette operators $B_p$ therefore have to be the same. Since $\partial A$ doesn't contain any non-contractible 2-cycle of the $T^3$, all the boundary conditions on $\partial A$ are related to each other, rendering the Schmidt coefficients $\lambda_{\{\tilde{a}_v\}}$ a constant. We can then obtain the entanglement entropy of the region $A$:
\begin{align}
S_{\psi,A} =  (|\partial A|-1) \log 2.
\end{align}
The same derivation applies to the $D^3$ region $B$ and results in $S_{\psi,B} =  (|\partial B|-1) \log 2$. For the region $A \cup B$, its two disconnected part $A$ and $B$ leads to two independent constraints on the permissible boundary conditions. The number of permissible boundary conditions on $\partial(A \cup B)$ is then given by $2^{|\partial (A\cup B)|-2}$. Again, since the Schmidt coefficients take a constant value, the entanglement entropy of $A\cup B$ is given by.
\begin{align}
S_{\psi,A \cup B} =  (|\partial (A\cup B)|-2) \log 2.
\end{align}
Notice that the $|\partial (A\cup B)|=|\partial A|+|\partial B|$. Since the 3+1D toric code model has 0 correlation length, we can directly obtain the LRMI $\mathcal{I}(D^3)$ as
\begin{align}
\mathcal{I}(D^3)=S_{\psi,A} +S_{\psi,B} -S_{\psi,A \cup B}=0,
\end{align}
which indicates the absence of CO in the 3+1D toric code model.

\subsection{Calculation of the LRMI $\mathcal{I}(D^2\times T^1_z)$}
We consider two separated $D^2\times T^1_z$ regions $A$ and $B$. The method of Schmidt decomposition of the ground state $|\psi \rangle$ is still valid. We first consider the region $A$. Due to the existence of a non-contractible 1-cycle, namely the non-contractible 1-cycle along the $z$ direction, in the region $A$, the Schmidt states are only labeled by the permissible boundary conditions $\{\tilde{a}_v\}$. They are also eigenstates of the WLO $W_z$ and, therefore, carries the labels $w_z=\pm 1$ which denotes their corresponding eigenvalues $w_z=\pm 1$. Therefore, we can write the Schmidt decomposition as
\begin{align}
|\psi\rangle= \sum_i \lambda_{\{\tilde{a}_v\},w_z} |\psi^A_{\{\tilde{a}_v\},w_z} \rangle |\psi^{\bar{A}}_{\{\tilde{a}_v\},w_z} \rangle
\end{align}
with the Schmidt states satisfying
\begin{align}
\tilde{A}_v|\psi^A_{\{\tilde{a}_v\},w_z} \rangle & = \tilde{a}_v |\psi^A_{\{\tilde{a}_v\},w_z} \rangle, ~~ \forall v \in \partial A
\nonumber \\
A_v|\psi^A_{\{\tilde{a}_v\},w_z} \rangle & = |\psi^A_{\{\tilde{a}_v\},w_z} \rangle, ~~ \forall v \in(A/\partial A)
\nonumber \\
B_p|\psi^A_{\{\tilde{a}_v\},w_z} \rangle & = |\psi^A_{\{\tilde{a}_v\},w_z} \rangle, ~~ \forall p \in  A
\nonumber \\
W_z|\psi^A_{\{\tilde{a}_v\},w_z} \rangle & = w_z |\psi^A_{\{\tilde{a}_v\},w_z}\rangle.
\end{align}
One can show, by matching the numbers of the equation above with the number of degrees of freedom contained in the region $A$,  these equation uniquely determine the Schmidt state $|\psi^A_{\{\tilde{a}_v\},w_z} \rangle$. Since the boundary $\partial A$ does not contain any non-contractible 2-cycles of the $T^3$, all the boundary conditions are related to each other, rendering the Schmidt coefficients only a function of their $w_z$ label, i.e. $\lambda_{\{\tilde{a}_v\},w_z}=\lambda_{w_z}$. The values of $\lambda_{w_z}$ is fixed (up to a $U(1)$ phase) by the expectation value of the WLO $W_z$ on the ground state $|\psi\rangle$: $\langle W_z \rangle_\psi =\frac{|\lambda_1|^2-|\lambda_{-1}|^2}{|\lambda_1|^2+|\lambda_{-1}|}$. Hence, we can obtain the entanglement entropy $S_{\psi,A}$ as
\begin{align}
S_{\psi,A}= (|\partial A|-1) \log 2 - \frac{1-\langle W_z \rangle_\psi}{2} \log \frac{1-\langle W_z \rangle_\psi}{2} -  \frac{1+\langle W_z \rangle_\psi}{2} \log \frac{1+\langle W_z \rangle_\psi}{2}.
\end{align}
The same derivation for the region $B$ shows that $S_{\psi,B}= (|\partial B|-1) \log 2 - \frac{1-\langle W_z \rangle_\psi}{2} \log \frac{1-\langle W_z \rangle_\psi}{2} -  \frac{1+\langle W_z \rangle_\psi}{2} \log \frac{1+\langle W_z \rangle_\psi}{2}$. For the region $A \cup B$, the Schmidt state are still labeled by their boundary conditions on $\partial(A \cup B)$ and the eigenvalue $w_z$ of the WLO $W_z$. Similar to the case of $\mathcal{I}(D^3)$, the number of permissible boundary conditions is again given by $2^{|\partial (A\cup B)|-2}$. The entanglement entropy $S_{\psi,A \cup B}$ is then given by
\begin{align}
S_{\psi,A \cup B} = (|\partial (A\cup B)|-2) \log 2 - \frac{1-\langle W_z \rangle_\psi}{2} \log \frac{1-\langle W_z \rangle_\psi}{2} -  \frac{1+\langle W_z \rangle_\psi}{2} \log \frac{1+\langle W_z \rangle_\psi}{2},
\end{align}
and the LRMI $\mathcal{I}(D^2\times T^1_z)$ is given by
\begin{align}
\mathcal{I}(D^2\times T^1_z) = S_{\psi,A}+S_{\psi,B} - S_{\psi,A \cup B}= - \frac{1-\langle W_z \rangle_\psi}{2} \log \frac{1-\langle W_z \rangle_\psi}{2} -  \frac{1+\langle W_z \rangle_\psi}{2} \log \frac{1+\langle W_z \rangle_\psi}{2},
\end{align}
which is the SE of the classical probability distribution of the state $|\psi\rangle$ on the two eigenvalues $w_z=\pm 1$ of the WLO $W_z$. As is proposed in the main text, this non-trivial value of $\mathcal{I}(D^2\times T^1_z)$ signifies the 1-membrane condensation in the 3+1D toric code model.

\subsection{Calculation of the LRMI $\mathcal{I}(D^1\times (T^2_{yz}/D^2))$}
We consider the two separated $D^1\times (T^2_{yz}/D^2)$ regions $A$ and $B$. Each of the regions $A$ and $B$ contains two non-contractible 1-cycles in the $y$ and $z$ directions along which the WLO's $W_y$ and $W_z$ can be constructed. Similar to the previous case, the Schmidt states rely not only on the boundary conditions, but also their eigenvalues $w_y, w_z=\pm $ with respect to the WLO's $W_y$ and $W_z$. Due to the non-existence of non-contractible 2-cycles of $T^3$ contained in $\partial A$, $\partial B$ or $\partial (A \cup B)$, the Schmidt coefficients do not depend on the boundary conditions, but only their $w_{y,z}$ labels. A calculation similar to the previous case leads the the conclusion that
\begin{align}
S_{\psi,A} & = (|\partial A|-1) \log 2 - \sum_{\gamma \in \{y,z\}}\left( \frac{1-\langle W_i \rangle_\psi}{2} \log \frac{1-\langle W_i \rangle_\psi}{2} +  \frac{1+\langle W_i \rangle_\psi}{2} \log \frac{1+\langle W_i \rangle_\psi}{2} \right)
\nonumber \\
S_{\psi,B} & = (|\partial B|-1) \log 2 - \sum_{\gamma \in \{y,z\}} \left( \frac{1-\langle W_i \rangle_\psi}{2} \log \frac{1-\langle W_i \rangle_\psi}{2} +  \frac{1+\langle W_i \rangle_\psi}{2} \log \frac{1+\langle W_i \rangle_\psi}{2} \right)
\nonumber \\
S_{\psi,A\cup B} & = (|\partial (A \cup B)|-2) \log 2 - \sum_{\gamma \in \{y,z\}} \left( \frac{1-\langle W_i \rangle_\psi}{2} \log \frac{1-\langle W_i \rangle_\psi}{2} +  \frac{1+\langle W_i \rangle_\psi}{2} \log \frac{1+\langle W_i \rangle_\psi}{2} \right).
\end{align}
The LRMI $\mathcal{I}(D^1\times (T^2_{xy}/D^2))$ is then given by
\begin{align}
\mathcal{I}(D^1\times (T^2_{xy}/D^2)) = S_{\psi,A} +S_{\psi,B} -S_{\psi,A \cup B} = - \sum_{\gamma \in \{y,z\}} \left( \frac{1-\langle W_i \rangle_\psi}{2} \log \frac{1-\langle W_i \rangle_\psi}{2} +  \frac{1+\langle W_i \rangle_\psi}{2} \log \frac{1+\langle W_i \rangle_\psi}{2} \right),
\end{align}
which coincides with the SE of the classical probability distribution of the state $|\psi\rangle$ on the 4 combination of eigenvalues $(w_x, w_y)$ of the two WLO's $W_{y,z}$.

\subsection{Calculation of the LRMI $\mathcal{I}(D^1\times T^2_{xy})$}
We consider the two separated $D^1\times T^2_{yz}$ regions $A$ and $B$. For the region $A$, the Schmidt decomposition of the ground state $\psi$, as we discussed in the previous case, is still given by
\begin{align}
|\psi\rangle= \sum_i \lambda_{\{\tilde{a}_v\},w_y,w_z} |\psi^A_{\{\tilde{a}_v\},w_y,w_z} \rangle |\psi^{\bar{A}}_{\{\tilde{a}_v\},w_y,w_z} \rangle
\end{align}
where $\{\tilde{a}_v\}$ denotes the boundary conditions and $w_{y,z}=\pm 1$ correspond to the eigenvalues of the WLO's $W_{y,z}$. In contrast to previous cases, due to the existence of non-contractible 2-cycles which is the one wrapping around the $y-z$ plane, all boundary conditions are not completely related to each other. To be more precise, the boundary $\partial A$ of the region $A$ consists of two disconnected parts $(\partial A)_{1,2}$, each with the topology of $T^2_{yz}$. The action of $B_p$ terms of any plaquette $p$ does not change the value of $\prod_{v \in (\partial A)_1} \tilde{a}_v$ (or equivalently $\prod_{v \in (\partial A)_2} \tilde{a}_v$). In fact, it is easy to show that $\prod_{v \in (\partial A)_1} \tilde{A}_v= W_{yz}$. Therefore, we can identify the $\prod_{v \in (\partial A)_1} \tilde{a}_v$ with the eigenvalue $w_{yz}$ of the 2-WSO $W_{yz}$, which can take values $\pm 1$. $w_{yz}$ provides a label for the topological sectors of the boundary conditions within which all the boundary conditions are related to each other. The Schmidt coefficients $\lambda_{\{\tilde{a}_v\},w_x, w_y}$ can then be rewritten as
\begin{align}
\lambda_{\{\tilde{a}_v\},w_y, w_z}=\lambda_{w_{yz} ,w_y, w_z} \Big|_{w_{yz} =\prod_{v \in (\partial A)_1} \tilde{a}_v},
\end{align}
which are then determined (up to a $U(1)$ phase) by the expectation values $\langle W_{yz} \rangle_\psi$, $\langle W_{y} \rangle_\psi$ and $\langle W_{z} \rangle_\psi$. With the values $w_{yz}$, $w_y$ and $w_z$ fixed, the number of permissible boundary conditions are given by $2^{|\partial A|-2}$. Therefore, the entanglement entropy $S_{\psi,A}$ is
\begin{align}
S_{\psi,A}= (|\partial A|-2) \log 2 - \sum_{\gamma \in \{yz, y,z\}}\left( \frac{1-\langle W_i \rangle_\psi}{2} \log \frac{1-\langle W_i \rangle_\psi}{2} +  \frac{1+\langle W_i \rangle_\psi}{2} \log \frac{1+\langle W_i \rangle_\psi}{2} \right).
\end{align}
Similarly the entanglement entropy $S_{\psi,B}$ and $S_{\psi,A \cup B}$ are given by
\begin{align}
S_{\psi,B} & = (|\partial B|-2) \log 2 - \sum_{\gamma \in \{yz, y,z\}} \left( \frac{1-\langle W_i \rangle_\psi}{2} \log \frac{1-\langle W_i \rangle_\psi}{2} +  \frac{1+\langle W_i \rangle_\psi}{2} \log \frac{1+\langle W_i \rangle_\psi}{2} \right)
\nonumber \\
S_{\psi,A\cup B} & = (|\partial (A \cup B)|-4) \log 2 - \sum_{\gamma \in \{yz, y,z\}} \left( \frac{1-\langle W_i \rangle_\psi}{2} \log \frac{1-\langle W_i \rangle_\psi}{2} +  \frac{1+\langle W_i \rangle_\psi}{2} \log \frac{1+\langle W_i \rangle_\psi}{2} \right).
\end{align}
Now, we obtain the LRMI $\mathcal{I}(D^1\times T^2_{yz})$ is then given by
\begin{align}
\mathcal{I}(D^1\times T^2_{xy}) = S_{\psi,A} +S_{\psi,B} -S_{\psi,A \cup B} = - \sum_{\gamma \in \{yz, y,z\}} \left( \frac{1-\langle W_i \rangle_\psi}{2} \log \frac{1-\langle W_i \rangle_\psi}{2} +  \frac{1+\langle W_i \rangle_\psi}{2} \log \frac{1+\langle W_i \rangle_\psi}{2} \right),
\end{align}
which coincide with the SE of the classical probability distribution of the state $|\psi\rangle$ on the 8 combination of eigenvalues $(w_{yz}, w_y, w_z)$ of the 2-WSO $W_{yz}$ and the two WLO's $W_{y,z}$. Generically, the LMRI $\mathcal{I}(D^1\times T^2_{yz}) $ differs from the LRMI $\mathcal{I}(D^1\times (T^2_{yz}/D^2))$. Therefore, we conclude that the 3+1D toric code model exhibits the 2-membrane condensation.

\section{Long-range mutual information in the 4+1D toric code models}
In this section, we will discuss the behavior of the LRMI in two different types of 4+1D toric code models\cite{Dennis2002_supp,Tarun2011_supp}. Using the LRMI diagnostics, we will show that one of them hosts coexisting 1-membrane and 3-membrane condensation and the other exhibits the 2-membrane condensation. To be more specific, we will consider the a generic ground state $|\psi\rangle$ in the ground state Hilbert space $\HG$ on the 4-torus $T^4$. The relevant LRMI here are $\mathcal{I}(D^4)$, $\mathcal{I}(D^3\times T^1_x)$, $\mathcal{I}(D^2\times T^2_{xy})$, $\mathcal{I}(D^2\times (T^2_{xy}/D^2))$, $\mathcal{I}(D^1\times T^3_{xyz})$ and $\mathcal{I}(D^1\times (T^3_{xyz}/D^3))$.

Let's first set up some conventions and terminology. The 4 spatial dimensions are denoted as $x$, $y$, $z$ and $u$. Both models are defined on a 4D hypercubic lattice. We will then refer to a vertex as a 0-cell, a link as a 1-cell, a 2D plaquette as a 2-cell, a 3D cube as a 3-cell and a 4D hypercube as a 4-cell. If an $m$-cell $c_m$ contains an $n$-cell $c_n$ ($n<m$), we interchangeably denote this relation as $c_m \in c_n$ or $c_n \in c_m$. For our LRMI diagnostic, we definite a region to be a collection of 4-cells such that if a 4-cell belongs to this region, so do all of its vertices,links, plaquettes and 3-cells. Also, if a 0-cell $v$, a 1-cell $l$, a 2-cell $p$ or a 3-cell $c$ belongs to this region, there must exist a 4-cells in the region such that this 4-cell contains $v$, $l$, $p$ or $c$ correspondingly. A vertex $v$ is defined to be on the boundary $\partial A$ of a region $A$ if $v \in A$ and there exists a link $l$ such that the link connects to the vertex $v$, namely $l \in v$, but $l \notin A$. A link $l$ is defined to be on the boundary $\partial A$ of a region $A$ if $l \in A$ and there exists a plaquette $p$ such that the plaquette contains the link $l$, namely $p \in l$, but $p \notin A$.

\subsection{4+1D toric code model with 1-membrane and 3-membrane condensation}
The 4+1D toric model that exhibits the 1-membrane and 3-membrane condensation is defined on the 4D hypercubic lattice with the degrees of freedom on the 1-cells (links) with the Hamiltonian given by
\begin{align}
H^\text{1,3-membrane}_\text{4DTC}= -\sum_{\text{0-cell}~v} A_v - \sum_{\text{2-cell}~p} B_p,
\end{align}
where the vertex operators $A_v$ and the plaquette operators $B_p$ are given by
\begin{align}
A_v= \prod_{\text{1-cell}~l \in v} \sigma_l^x, ~~~ B_p = \prod_{\text{1-cell}~l \in p} \sigma_l^z.
\end{align}
All the terms in $H^\text{1,3-membrane}_\text{4DTC}$ commute with each other, rendering this model exactly solvable. The ground state of  $H^\text{1,3-membrane}_\text{4DTC}$ on $S^4$ is non-degenerate. However, there is $2^4$-fold ground state degeneracy on the 4-torus $T^4$, which is a signature of topological order. Within the ground state Hilbert space $\HG$ on the $T^4$, WLO's, including $W_{x}$, $W_{y}$, $W_{z}$ and $W_{u}$, and the 3-WSO's, including $W_{xyz}$, $W_{yzu}$, $W_{zux}$ and $W_{uxy}$, act non-trivially. The subscripts of the WLO's and the 3-WSO's represent the non-contractible prime 1-cycles or 3-cycles of the $T^4$ these operators are constructed along. Each of them has eigenvalues $\pm 1$. All of the non-trivial commutation relations among them are given by
\begin{align}
W_x W_{yzu} = - W_{yzu} W_x, ~~~~  W_y W_{zux} = - W_{zux} W_y, ~~~~  W_z W_{uxy} = - W_{uxy} W_z, \text{ and }  W_u W_{xyz} = - W_{xyz} W_u.
\end{align}

Now we present a general discussion for the region of $T^4$ with all possible topologies relevant to the LRMI diagnostics. On a surbregion $A$, we can define, for every boundary vertex $v\in \partial A$, a boundary vertex operator $\tilde{A}_v= \prod_{\text{1-cell}~l, ~ l\in v ~\&~l \in A } \sigma_l^x $. A choice of eigenvalues $\tilde{a}_v=\pm 1$ for each of the boundary vertex operator $\tilde{A}_v$ will be referred to as a boundary condition $\{\tilde{a}_v\}_{v\in \partial A}$ on $\partial A$. For all the relevant cases, the following Schmidt decomposition applies:
\begin{align}
|\psi\rangle= \sum_i \lambda_{\{\tilde{a}_v\},\{w_{\gamma_1}\}} |\psi^A_{\{\tilde{a}_v\},\{w_{\gamma_1}\}} \rangle |\psi^{\bar{A}}_{\{\tilde{a}_v\},\{w_{\gamma_1}\}} \rangle,
\end{align}
where $\{w_{\gamma_1}\}$ is a set of choices of the eigenvalues $w_{\gamma_1}=\pm 1$ for all the WLO's $W_{\gamma_1}$ contained in the region $A$. The boundary conditions are all related to each other unless non-trivial topological sectors exists. For every non-contractible 3-cycles $\gamma_3$ of the $T^4$ contained in the region $A$, there is a label $w_{\gamma_3}=\pm$ for the topological sectors of the boundary conditions. This label $w_{\gamma_3}$ correspond to a choice of the eigenvalue of the associated 3-WSO $W_{\gamma_3}$. Only the boundary conditions with the same topological sector are related to each other. Therefore, the values of Schmidt coefficients only depends on the labels $\{w_{\gamma_3}\}$ and the $\{w_{\gamma_1}\}$. When the non-contractible 1-cycles or 3-cycles does not exist, the corresponding labels drop out automatically. Also, for all the relevant cases, we notice that the numbers of permissible boundary conditions satisfy that if the region $A$ is a union of two disjoint regions (say $A^1$ and $A^2$), the number of permissible boundary conditions on $\partial A$ is a product of the number of permissible boundary conditions on $\partial A^1$ and that on $\partial A^2$. With these general understanding of the Schmidt coefficients, we can derive the all the relevant LRMI. The result are summarized as follows:
\begin{align}
&\mathcal{I}(D^4)  = 0,
\nonumber \\
&\mathcal{I}(D^3\times T^1_x)  = - \sum_{\gamma \in\{x\}} \left( \frac{1-\langle W_i \rangle_\psi}{2} \log \frac{1-\langle W_i \rangle_\psi}{2} +  \frac{1+\langle W_i \rangle_\psi}{2} \log \frac{1+\langle W_i \rangle_\psi}{2} \right),
\nonumber \\
&\mathcal{I}(D^2\times T^2_{xy})  = - \sum_{\gamma \in \{x, y\}} \left( \frac{1-\langle W_i \rangle_\psi}{2} \log \frac{1-\langle W_i \rangle_\psi}{2} +  \frac{1+\langle W_i \rangle_\psi}{2} \log \frac{1+\langle W_i \rangle_\psi}{2} \right),
\nonumber \\
&\mathcal{I}(D^2\times  (T^2_{xy}/D^2))  = - \sum_{\gamma \in \{x, y\}} \left( \frac{1-\langle W_i \rangle_\psi}{2} \log \frac{1-\langle W_i \rangle_\psi}{2} +  \frac{1+\langle W_i \rangle_\psi}{2} \log \frac{1+\langle W_i \rangle_\psi}{2} \right),
\nonumber \\
&\mathcal{I}(D^1\times T^3_{xyz})  = - \sum_{\gamma \in \{x, y, z, xyz \}} \left( \frac{1-\langle W_i \rangle_\psi}{2} \log \frac{1-\langle W_i \rangle_\psi}{2} +  \frac{1+\langle W_i \rangle_\psi}{2} \log \frac{1+\langle W_i \rangle_\psi}{2} \right),
\nonumber \\
&\mathcal{I}(D^1\times (T^3_{xyz}/D^3))  = - \sum_{\gamma \in \{x, y, z \}} \left( \frac{1-\langle W_i \rangle_\psi}{2} \log \frac{1-\langle W_i \rangle_\psi}{2} +  \frac{1+\langle W_i \rangle_\psi}{2} \log \frac{1+\langle W_i \rangle_\psi}{2} \right).
\end{align}
The generically non-vanishing values of $\mathcal{I}(D^3\times T^1_x) - \mathcal{I}(D^4) $ and $\mathcal{I}(D^1\times T^3_{xyz}) - \mathcal{I}(D^1\times (T^3_{xyz}/D^3))$ signify the existence of 1-membrane and 3-membrane condensations. In contrast, the fact that $0=\mathcal{I}(D^4) = \mathcal{I}(D^2\times T^2_{xy}) - \mathcal{I}(D^2\times  (T^2_{xy}/D^2)) $ indicates the absence of the 0-membrane and 2-membrane condensations.

\subsection{4+1D toric code model with 2-membrane condensation}
The 4+1D toric model that exhibits the 2-membrane and 3-membrane condensation is defined on the 4D hypercubic lattice with the degrees of freedom on the 2-cells (plaquettes) with the Hamiltonian given by
\begin{align}
H^\text{2-membrane}_\text{4DTC}= -\sum_{\text{1-cell}~l} A'_l - \sum_{\text{3-cell}~c} B'_c ,
\end{align}
where the link operators $A'_l$ and the cube operators $B'_c$ are given by
\begin{align}
A'_l= \prod_{\text{2-cell}~p \in l} \sigma_p^x, ~~~  B_c = \prod_{\text{2-cell}~p \in c} \sigma_p^z.
\end{align}
This model is exactly sovable, since All the terms in $H^\text{2-membrane}_\text{4DTC}$ commute with each other. The ground state of $H^\text{2-membrane}_\text{4DTC}$ is non-degenerate. However, there is $2^6$-fold ground state degeneracy on the 4-torus $T^4$, which is a signature of topological order. Within the ground state Hilbert space $\HG$ on the $T^4$, the 2-WSO's, $W^{\alpha}_{xy}$, $W^{\alpha}_{yz}$, $W^{\alpha}_{zx}$, $W^{\alpha}_{xu}$, $W^{\alpha}_{yu}$ and $W^{\alpha}_{zu}$ ($\alpha\in\{e,m\}$), act non-trivially. The subscripts of the 2-WSO's represents the non-contractible prime 2-cycles they are constructed along. The superscripts $e$ and $m$ represent the electric WSO's and the magnetic WSO's respectively. Each of them has eigenvalues $\pm 1$. All of the non-trivial commutation relations among them are given by
\begin{align}
W^{e}_{xy} W^{m}_{zu}= -  W^{m}_{zu} W^{e}_{xy}, ~~W^{e}_{yz} W^{m}_{xu}= -  W^{m}_{xu} W^{e}_{yz}, ~~W^{e}_{zx} W^{m}_{yu}= -  W^{m}_{yu} W^{e}_{zx},
\nonumber \\
W^{e}_{xu} W^{m}_{yz}= -  W^{m}_{yz} W^{e}_{xu}, ~~W^{e}_{yu} W^{m}_{zx}= -  W^{m}_{zx} W^{e}_{yu}, ~~ W^{e}_{zu} W^{m}_{xy}= -  W^{m}_{xy} W^{e}_{zu}.
\end{align}

Similar to the previous case, we will present a general discussion for the region of $T^4$ with all possible topologies relevant to the LRMI diagnostics. On a region $A$, for every boundary link $l\in \partial A$, we can define the boundary link operator $\tilde{A}'_l= \prod_{\text{2-cell}~p, ~ p\in l ~\&~p \in A } \sigma_p^x $. A choice of eigenvalues $\tilde{a}'_l=\pm 1$ for each of the boundary link operator $\tilde{A}'_l$ will be referred to as a boundary condition $\{\tilde{a}'_l\}_{l\in \partial A}$ on $\partial A$ for this model. For all the relevant cases, the following Schmidt decomposition applies:
\begin{align}
|\psi\rangle= \sum_i \lambda_{\{\tilde{a}'_l\},\{w^m_{\gamma_2}\}} |\psi^A_{\{\tilde{a}'_v\},\{w^m_{\gamma_2}\}} \rangle |\psi^{\bar{A}}_{\{\tilde{a}'_v\},\{w^m_{\gamma_2}\}} \rangle,
\end{align}
where $\{w^m_{\gamma_2}\}$ is a set of choices of the eigenvalues $w^m_{\gamma_2}=\pm 1$ for all 2-WSO's $W^m_{\gamma_2}$, if exist, contained in the region $A$. Here $\gamma_2$ labels the non-contractible 2-cycles contained in the region $A$ along which the 2-WSO $W^m_{\gamma_2}$ is constructed. In the presence of non-contractible 2-cycles, the boundary conditions are also divided into different different topological sector whose label is given by the set choices $\{w^e_{\gamma_2} \}$ for the eigenvalues of all the 2-WSO's  $W^e_{\gamma_2}$ contained in the region $A$. Within each topological sector, the boundary conditions are related to each other, rendering the Schmidt coefficient only a function of $\{w^e_{\gamma_2}\}$ and the $\{w^m_{\gamma_2}\}$. In the absence of non-contractible 2-cycles, the corresponding labels drop out automatically. Also, in all the relevant cases, the numbers of permissible boundary conditions satisfy that if the region $A$ is a union of two disjoint regions (say $A^1$ and $A^2$), the number of permissible boundary conditions on $\partial A$ is a product of the number of permissible boundary conditions on $\partial A^1$ and that on $\partial A^2$. With these general understanding of the Schmidt coefficients, we can derive the all the relevant LRMI. The results are summarized as follows:
\begin{align}
&\mathcal{I}(D^4)  = 0,
\nonumber \\
&\mathcal{I}(D^3\times T^1_x)  = 0,
\nonumber \\
&\mathcal{I}(D^2\times T^2_{xy})  = - \sum_{\alpha \in \{e,m\}} \left( \frac{1-\langle W^\alpha_{xy} \rangle_\psi}{2} \log \frac{1-\langle W^\alpha_{xy} \rangle_\psi}{2} +  \frac{1+\langle W^\alpha_{xy} \rangle_\psi}{2} \log \frac{1+\langle W^\alpha_{xy} \rangle_\psi}{2} \right),
\nonumber \\
&\mathcal{I}(D^2\times  (T^2_{xy}/D^2))  = 0,
\nonumber \\
&\mathcal{I}(D^1\times T^3_{xyz})  = - \sum_{\alpha \in \{e,m\}}\sum_{\gamma \in \{xy, yz, zx \}} \left( \frac{1-\langle W^\alpha_i \rangle_\psi}{2} \log \frac{1-\langle W^\alpha_i \rangle_\psi}{2} +  \frac{1+\langle W^\alpha_i \rangle_\psi}{2} \log \frac{1+\langle W^\alpha_i \rangle_\psi}{2} \right),
\nonumber \\
&\mathcal{I}(D^1\times (T^3_{xyz}/D^3))  = - \sum_{\alpha \in \{e,m\}} \sum_{\gamma \in \{xy, yz, zx \}} \left( \frac{1-\langle W^\alpha_i \rangle_\psi}{2} \log \frac{1-\langle W^\alpha_i \rangle_\psi}{2} +  \frac{1+\langle W^\alpha_i \rangle_\psi}{2} \log \frac{1+\langle W^\alpha_i \rangle_\psi}{2} \right).
\end{align}
The generically non-vanishing values of $ \mathcal{I}(D^2\times T^2_{xy}) - \mathcal{I}(D^2\times  (T^2_{xy}/D^2)) $ signifies the existence of the 2-membrane condensation. In contrast, the fact that $0=\mathcal{I}(D^4) =\mathcal{I}(D^3\times T^1_x)   =\mathcal{I}(D^1\times T^3_{xyz}) - \mathcal{I}(D^1\times (T^3_{xyz}/D^3))$ indicates the absence of the 0-membrane, 1-membrane and 3-membrane condensations.

\end{document}